\documentclass[review]{elsarticle}

\usepackage{amsmath,amssymb} 
\usepackage[section]{algorithm} 
\usepackage{algorithmic}
\usepackage{epsfig}
\usepackage{acronym}
\usepackage{subfigure}
\usepackage{bm}
\usepackage{multirow}
\usepackage{bbding}

\begin{document}

\begin{frontmatter}

\acrodef{N$^3$SC}{Normalized Non-Negative Sparse Coding}
\acrodef{NNSC}{Non-Negative Sparse Coding}
\acrodef{NMF}{Non-negative Matrix Factorization}

\title{PC-GANs: Progressive Compensation Generative Adversarial Networks for Pan-sharpening}

\author[mymainaddress1,mymainaddress3]{Yinghui Xing}

\author[mymainaddress2]{Shuyuan Yang\corref{mycorrespondingauthor}}
\cortext[mycorrespondingauthor]{Corresponding author.}
\ead{syyang2009@gmail.com}

\author[mymainaddress1,mymainaddress3]{Song Wang}
\author[mymainaddress1,mymainaddress3]{Yan Zhang}
\author[mymainaddress1,mymainaddress3]{Yanning Zhang}
\address[mymainaddress1]{Integrated Aerospace-Ground-Ocean Big Data Application Technology, Northwestern Polytechnical University, Xi’an, Shaanxi, China 710072}
\address[mymainaddress2]{School of Artificial Intelligence, Xidian University, Xi'an, Shaanxi, China, 710071}
\address[mymainaddress3]{School of Computer Science, Northwestern Polytechnical University, Xi’an, Shaanxi, China 710072}

\begin{abstract}
The fusion of multispectral and panchromatic images is always dubbed pan-sharpening. Most of the available deep learning-based pan-sharpening methods sharpen the multispectral images through a one-step scheme, which strongly depends on the reconstruction ability of the network. However, remote sensing images always have large variations, as a result, these one-step methods are vulnerable to the error accumulation and thus incapable of preserving spatial details as well as the spectral information. In this paper, we propose a novel two-step model for pan-sharpening that sharpens the MS image through the progressive compensation of the spatial and spectral information. Firstly, a deep multiscale guided generative adversarial network is used to preliminarily enhance the spatial resolution of the MS image. Starting from the pre-sharpened MS image in the coarse domain, our approach then progressively refines the spatial and spectral residuals over a couple of generative adversarial networks (GANs) that have reverse architectures. The whole model is composed of triple GANs, and based on the specific architecture, a joint compensation loss function is designed to enable the triple GANs to be trained simultaneously. Moreover, the spatial-spectral residual compensation structure proposed in this paper can be extended to other pan-sharpening methods to further enhance their fusion results. Extensive experiments are performed on different datasets and the results demonstrate the effectiveness and efficiency of our proposed method.
\end{abstract}


\end{frontmatter}


\acresetall

\section{Introduction}\label{sec_Intro}
Pan-sharpening aims to enhance the resolution of multispectral (MS) image with the help of a panchromatic (PAN) image. MS images have high spectral resolution but low spatial resolution, while PAN images have high spatial resolution and low spectral resolution. Then pan-sharpening is also an image fusion technique, to obtain an image that has high-resolution in both spatial and spectral domain. Such an image has a wide range of applications, such as image classification \cite{shu2015object,yang2018self}, change detection \cite{liu2016deep}, \cite{zhang2018coarse} and segmentation \cite{ghamisi2013multilevel}. Compared with natural image fusion, remote sensing (RS) image fusion has its specific complexity due to the fact that RS images always have a wider horizon and eventually contain more types of land covers. Furthermore, the abundant spectral bands of MS images also exacerbate the difficulties. The specific characteristics of RS images and the rigorous demand that an MS image should have both high spatial and high spectral resolution for subsequent applications increase the difficulties of pan-sharpening task.

Over the last decade, various pan-sharpening methods are proposed \cite{vivone2014critical,baronti2011theoretical,yin2017joint}. They can be divided into three categories: component substitution (CS) based methods, multiresolution analysis (MRA) based methods and optimization restoration (OR) based methods \cite{yin2017joint}. CS family first projects MS channels to other spaces where the spatial and spectral components are assumed to be separated, then the spatial component of MS is replaced with the PAN image.  Improved adaptive intensity-hue-saturation (IAIHS) method \cite{leung2013improved}, nonlinear intensity-hue-saturation (NIHS) \cite{ghahremani2016nonlinear} method and clustered image-based method \cite{shahdoosti2017pansharpening} are examples of recently proposed CS-based methods. Because the overlap between spatial and spectral components exists more or less, the substitution of the spatial component results in serious spectral distortion. MRA family extracts the spatial details of the PAN image through a multiresolution decomposition such as contourlet transform \cite{yang2010image}, curvelet transform \cite{nencini2007remote}, support tensor transform \cite{xing2018pansharpening}, and morphological operators \cite{restaino2016fusion}, then the spatial details are injected into the low-resolution MS image. This type of methods can better reduce the spectral distortion generally, but they tend to produce some spatial degradation like blocky and aliasing artifacts. OR-based methods mainly recast pan-sharpening as an inverse problem and try to restore the high-resolution MS image from PAN and low-resolution MS image. Owing to that the restoration is an ill-posed problem, total variation \cite{vivone2014pansharpening} and sparse representation \cite{he2014new} are usually employed as the regularization terms for OR-based methods. This type of methods proved to be effective not only in preserving spectral information but also in improving the spatial details of original MS images.

Recently, deep learning (DL) based methods have been proposed to pan-sharpening \cite{huang2015new,xing2018pan,masi2016pansharpening,wei2017boosting,yang2017pannet,yuan2018multiscale}. The first DL based pan-sharpening method assumed that the relationship between high-resolution and low-resolution MS image patches is the same as that between the corresponding PAN image patches, and then utilized a sparse denoising auto-encoder to learn this relationship. Similarly, Xing \emph{et al}. \cite{xing2018pan} also utilized the auto-encoder model to reconstruct high-resolution MS patches based on low/high-resolution PAN image patches and low-resolution MS image patches, but the authors first divided image patches into several categories to construct different geometric manifolds and then trained the auto-encoders with corresponding image patches. Motivated by the super-resolution convolutional neural network (SRCNN) \cite{dong2015image} that is proposed for natural image super-resolution task, Masi \emph{et al}. \cite{masi2016pansharpening} proposed a pan-sharpening neural network (PNN), in which the convolutional neural network (CNN) is used to model the pan-sharpening process as an end-to-end mapping. Although the architecture of PNN is the same as SRCNN that is designed for natural images, the domain-specific knowledge in RS imagery is introduced to improve the fusion performance. Later, Scarpa \emph{et al}. \cite{scarpa2018target} explored several architectural and training variations to the PNN baseline, and achieved further performance gains with a lightweight network that trains very fast. Following the idea of \cite{masi2016pansharpening}, Wei \emph{et al}. introduced residual learning into CNN based pan-sharpening \cite{wei2017boosting}, where a much deeper network is employed to take full advantage of residual learning. Supported by the residual learning architecture, the network can be designed deep, and the training can converge quickly. However, these methods simply regard pan-sharpening as a regression problem and do not emphasize the spatial enhancement and spectral preservation \cite{yang2017pannet}. To take problem-specific knowledge into consideration, Yang \emph{et al}. utilized a “spectra-mapping” procedure to propagate spectral information \cite{yang2017pannet}, and the model is trained in the high-pass domain to focus on spatial structures. Yuan \emph{et al}. also proposed a multiscale and multi-depth convolutional neural network (MSDCNN) \cite{yuan2018multiscale} that made use of three different sizes of convolutional filters to achieve multiscale feature extraction and two networks with different depth to achieve multi-depth. The authors believed that multiscale feature extraction helps to learn more robust convolutional filters.

Motivated by the facts that GANs can provide a powerful framework for generating plausible-looking natural images and that the adversarial procedure encourages the reconstructions to move towards regions of the search space with a high probability of containing photo-realistic images \cite{ledig2017photo}, in this paper, we propose a two-step pan-sharpening framework based on progressive compensation generative adversarial networks (PC-GANs). The PC-GANs model is mainly composed of two modules, i.e., deep multiscale guidance (DMG) module and the spatial-spectral residual compensation (SSRC) module. In the first step, the PAN features are utilized as the deep spatial guidance to enhance the spatial resolution of MS image, which is in accordance with the main propose of pan-sharpening, and the guidance network can be trained by strong supervision. In order to further emphasize the spatial details and spectral information of the pre-pansharpened MS image, a spatial-spectral residual compensation module is used. The SSRC module is composed of two reverse generative adversarial networks (GANs) that propagate spatial-spectral information from coarse-to-fine (C2F) or fine-to-coarse (F2C), which ensures the information flow cyclically. Finally, based on the structure of PC-GANs, a new loss function is designed to train the triple GANs simultaneously. In general, the main contributions are as follows:

1) We design a two-step pan-sharpening method by progressively compensating the spatial and spectral residuals on the pre-sharpened MS image. The compensation in both the coarse and fine domain helps to automatically refine the result of the first step, and well enhance the spatial and spectral resolution of it. 

2) An SSRC module that consists of a couple of reverse GANs is proposed to map the spatial-spectral residuals cyclically. Such a structure makes the compensation of spatial-spectral residuals realized in both the coarse and fine domains by a coupled GANs and enhances the spatial details and at the same time improves the spectral resolution.

3) For the unique structure of PC-GANs, we propose a joint compensation loss function that enables the triple GANs to be trained simultaneously.

The remainder of this paper is organized as follows. Section \ref{sec_RW} analyzes the related GAN-based pan-sharpening methods. Section \ref{sec_ME} details our proposed approach. Experimental results and corresponding discussions are presented in Section \ref{sec_EXP}. In Section \ref{sec_ABL}, we provide the ablation studies. The analysis of SSRC strategy is given in Section Section \ref{sec_ASS}. Finally, conclusions are given in Section \ref{sec_CON}.

\begin{figure}[h]
	\centering
	\subfigure[]{\includegraphics[scale=1.5]{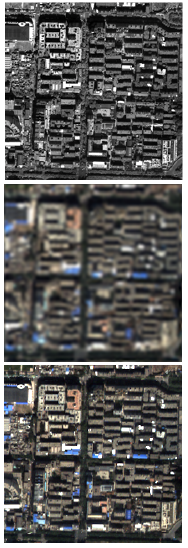}}
	\subfigure[]{\includegraphics[scale=1.5]{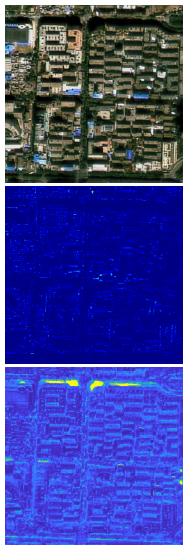}}
	\subfigure[]{\includegraphics[scale=1.5]{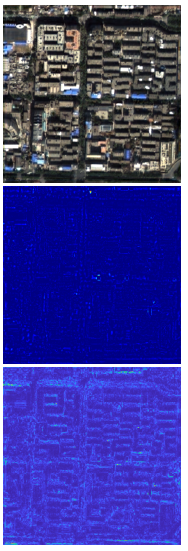}}
	\subfigure{\includegraphics[scale=1.5]{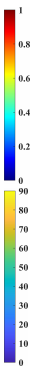}}
	\caption{Fusion results of one-step GAN-based method (PanGAN) and proposed method. (a) Source images and the reference image. (b) PanGAN \cite{ma2020pan}, and (c) Proposed method. The first row, second row and the third row of (b) and (c) show the fusion results, the difference maps and the SAM maps, respectively. }\label{fig1}
\end{figure}

\section{Related Work}\label{sec_RW}
The first pan-sharpening method that utilized the GAN structure is the Pan-sharpening GAN (PSGAN) \cite{liu2020psgan}. In PSGAN, a two-stream fusion architecture acted as a generator, and a fully convolutional network acted as the discriminator. However, due to the fact that the fusion is conducted by encoding the concatenated features extracted from MS and PAN, the importance of PAN images is not prominent. Zhu \emph{et al}. \cite{zhu2020super} proposed a multi-discriminator GAN structure that discriminated not only the pixel differences but also two major attributes in pan-sharpening, i.e., spectral preservation and the spatial enhancement. But the generator is modeled by a pre-trained VGG network, which is not designated for the remote sensing imagery, and leads to spectral distortion. Later, Gastineau \emph{et al}. \cite{gastineau2021generative} transformed original images to YCbCr space and then utilized two discriminators to decouple the two tasks of spectral preservation and spatial preservation. Though it could obtain a better fusion result, the transformation of MS to YCbCr space is still controversial. Similarly, Ma \emph{et al}. \cite{ma2020pan} also proposed an unsupervised pan-sharpening framework, i.e., PanGAN, where the generator was associated with a spectral discriminator and a spatial discriminator. PanGAN is totally an unsupervised pan-sharpening method and obtains well results. All above GAN-based pan-sharpening methods are one-step fusion, which strongly depends on the reconstruction ability of the network. However, remote sensing images always have large variations, as a result, fusion models should be carefully designed and trained to make them keep spatial and spectral details as many as possible. Figure \ref{fig1} shows the spatial difference and spectral angle mapper (SAM) of the one-step GAN-based pan-sharpening method and the proposed one, where the higher values of SAM means more serious spectral distortion. It can be observed that the one-step GAN-based method loss some spatial details, and has higher SAM values. Because the SAM map can well measure the loss of spectral information, it means that the spectral distortion also happens.

\begin{figure}[t]
	\centering
		\includegraphics[width=0.8\textwidth]{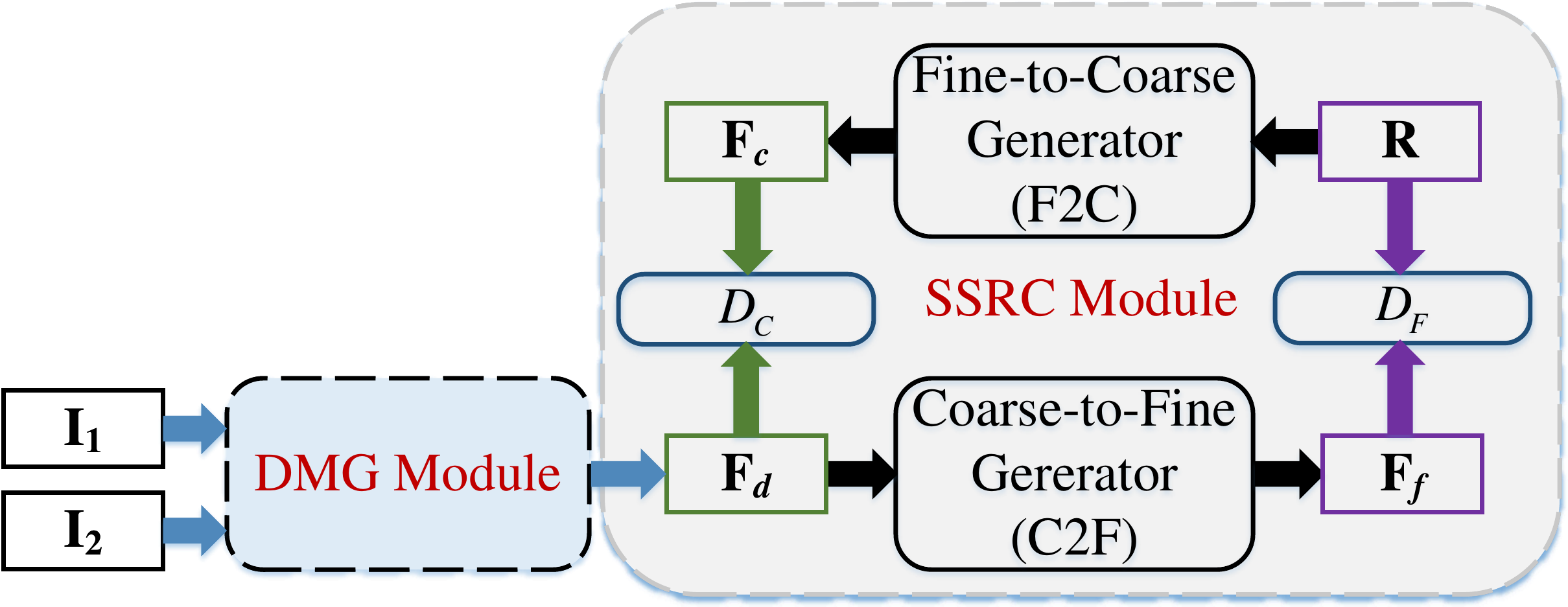}
	  \caption{Structure of PC-GANs, where it is composed of a deep multiscale guidance (DMG) module and a spatial-spectral residual compensation (SSRC) module.}\label{fig2}
\end{figure}

\section{Proposed Method}\label{sec_ME}

The whole PC-GANs model shown in Figure \ref{fig2} mainly contains two modules, i.e., DMG module and SSRC module. In this section, we first provide an overview of the PC-GANs, and then detail the two modules. Finally, aiming at this specific structure, a joint compensation loss function is proposed to train this model efficiently and thoroughly.

\subsection{Overview of the PC-GANs}\label{subsec_pcgans}
Figure \ref{fig2} shows the overall structure of proposed method. PC-GANs have two modules, i.e., DMG module and SSRC module, where the SSRC module is composed of a C2F generator, an F2C generator and two discriminators $D_C$ and $D_F$.

In general, PC-GANs take the PAN image $\mathbf{I}_1 \in \mathbb{R}^{M \times M}$  and the MS image $\mathbf{I}_2 \in \mathbb{R}^{m \times m \times b}$ as inputs, where $M=r \times m$ and $r$ is the ratio of their spatial resolution.In the first step, they are prefused by DMG module, and the pre-fused image $\mathbf{F}_d \in \mathbb{R}^{M \times M \times b}$ is then taken as the input of C2F to generate the final fusion result $\mathbf{F}_f \in \mathbb{R}^{M \times M \times b}$ of which the spatial-spectral information can be compensated by the SSRC module. Correspondingly, an F2C takes the reference image $\mathbf{R} \in \mathbb{R}^{M \times M \times b}$ as the input and generates an adaptive image $\mathbf{F}_c \in \mathbb{R}^{M \times M \times b}$ in the coarse domain to accomplish the refinement of the result of DMG module. The inputs of $D_F$ are $\mathbf{F}_f$ and $\mathbf{R}$,both of which are in the fine domain, therefore,$D_F$ is the discriminator in the fine domain. On the contrary,$\mathbf{F}_c$ and $\mathbf{F}_d$ are in the coarse domain, and they are taken as the inputs of the discriminator $D_C$ in the coarse domain.

C2F and $D_F$ together with F2C and $D_C$ constitute two GANs. Actually, the DMG module and $D_C$ also form a GAN structure. The main difference between $\{ F2C,D_C\}$ and $\{DMG,D_C\}$ pairs is that the real samples for $\{DMG,D_C\}$ pair are fake samples for $\{ F2C,D_C\}$ pair. That is, for example, $\mathbf{F}_c$ is the real sample for $\{DMG,D_C\}$ pair, however, it is the generated fake sample for $\{ F2C,D_C\}$ pair. As a result, $\{ F2C,D_C\}$, $\{DMG,D_C\}$ and $\{C2F,D_F\}$ constitute our multiscale progressive compensation GANs (PC-GANs) model. The PC-GANs model continues to be optimized cyclically until reaches an equilibrium, and the spatial-spectral residuals are progressively compensated in this process.

\begin{figure}[h]
	\centering
	\includegraphics[width=0.8\textwidth]{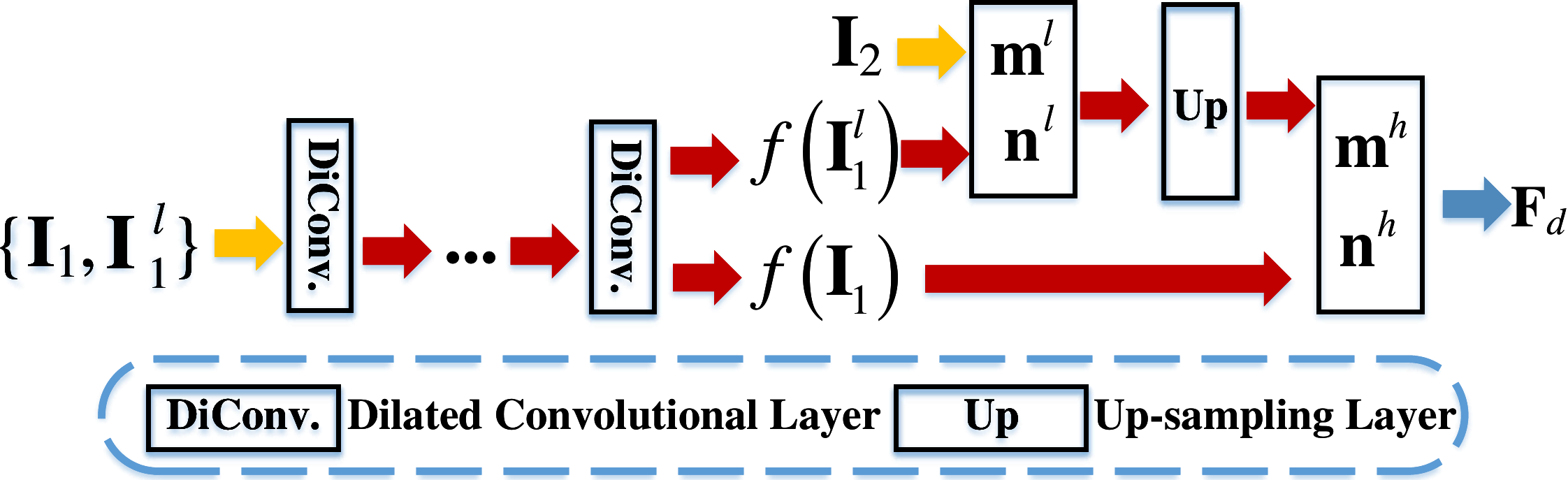}
	\caption{Structure of DMG module, where the low-resolution coefficients $\{\mathbf{m}^l, \mathbf{n}^l\}$ are computed from available low-resolution images while high-resolution coefficients $\{\mathbf{m}^h, \mathbf{n}^h\}$ are computed by up-sampling the low-resolution coefficients.}\label{fig3}
\end{figure}

\subsection{Deep Multiscale Guidance Module}\label{subsec_dmgm_extra}
Our DMG module is motivated by the guided filter, which is first proposed in \cite{he2012guided} and has wide applications as its edge-preserving property. In guided image filtering, the input image is filtered with the help of a guidance map, and the guidance map can be the input image itself or another different image. The basic assumption of the guided filter is a local linear model between the guidance map and the filtering output. In this paper, we propose a deep multiscale guidance model based on the process of guided filtering that maps the PAN image from pixel space to feature space to form a deep guidance map. The DMG module uses the dilated convolution to aggregate the multiscale information contained in PAN and keeps the spatial information to a great extent. Finally, the direct supervision which benefits a lot for spectral information preservation is utilized to preserve the spectral information as much as possible \cite{ma2020pan}.

We first obtain the down-sampled version of $\mathbf{I}_1$ by
\begin{equation}
\mathbf{I}_1^l=d_r(\mathbf{I}_1)
\end{equation}
where $d_r(\cdot)$ is the down-sampling operation that down-samples the image by a factor of $r$. Then $\mathbf{I}_1$ and ${\mathbf{I}_1}^l$ are fed to the network $f(\cdot)$ to map the inputs to the feature space where the MS image $\mathbf{I}_2$ can be linearly represented by the deep guidance maps, i.e.,
\begin{equation}
\mathbf{I}_{2,j}=m_k^lf(\mathbf{I}_{1,j}^l)+n_k^l, \forall j \in w_k,
\end{equation}
where $w_k$ is a window that centered at the pixel $k$, and $(m_k^l, n_k^l)$ are linear coefficients between the two low spatial resolution images, and they are constant in $w_k$.The coefficients can be estimated by minimizing the squared difference
\begin{equation}
E(m_k^l, n_k^l)= \sum_{j \in w_k}\big({(m_k^lf(\mathbf{I}_{1,j}^l)+n_k^l-\mathbf{I}_{2,j})}^2+\lambda{(m_k^l)}^2\big),
\end{equation}
in which $\lambda$ is a regularization parameter. Because (3) is a linear ridge regression model \cite{hastie2009elements}, its solution is given by
\begin{equation}
m_k^l=\frac{\frac{1}{\lvert w \lvert}\sum_{j \in w_k} f(\mathbf{I}_{1,j}^l)\mathbf{I}_{2,j}-\mu_k{\bar{x}}_k}{\sigma_k^2+\lambda}
\end{equation}
\begin{equation*}
n_k^l={\bar{x}}_k-m_k^l\mu_k,
\end{equation*}
where $\lvert w \lvert$ denotes the number of pixels contained in $w_k$ and ${\bar{x}}_k=\frac{1}{\lvert w \lvert} \sum_{j \in w_k} \mathbf{I}_{2,j}$ is the mean value of $\mathbf{I}_{2,j}$ in $w_k$. $\mu_k$ and ${\sigma_k}^2$ are the mean and variance of the guidance $f(\mathbf{I}_{1,j}^l)$ in $w_k$. After obtaining $(m_k^l, n_k^l)$, we can obtain the low-resolution representation coefficients $\mathbf{m}^l$ and $\mathbf{n}^l$. Afterwards, the high-resolution representation coefficients $\mathbf{m}^h$ and $\mathbf{n}^h$ are obtained by
\begin{equation}
\begin{split}
&\mathbf{m}^h=u_r(\mathbf{m}^l) \\
&\mathbf{n}^h=u_r(\mathbf{n}^l)
\end{split}
\end{equation}
where $u_r(\cdot)$ is the up-sampling operation and the up-sampling factor is $r$. Correspondingly, the fusion result $\mathbf{F}_d$ becomes 
\begin{equation}
\mathbf{F_d}=\mathbf{m}^h \otimes f(\mathbf{I}_1) + \mathbf{n}^h,
\end{equation}
where ''$\otimes$'' represents the element-wised multiplication.

In this paper, the nonlinear mapping function $f(\cdot)$ is modeled by a series of dilated convolutional layers. Because the dilated convolution can aggregate long-range contextual information without losing resolution \cite{chen2017fast}, it ensures the preservation of spatial details in a multiscale manner. Let $O^l (l=1,...,L)$ represent the output of $l^{th}$ layer in the network, then
\begin{equation}
O^l=\Phi(\mathbf{w}_{l-1}*_\gamma O^{l-1}+b_{l-1})
\end{equation}
where $\mathbf{w}_{l-1}$ and $b_{l-1}$ are the weights and biases of the network, and $\Phi(\cdot)$ is the nonlinear activation function. The symbol ''$*_\gamma$'' is a dilated convolution with a dilation rate $\gamma$. In this paper, the dilation rate is increased exponentially for the layers $3 \le l \le L-2$ and $\gamma=1$ for the first two layers together with the last two layers.
\par
The training loss for the DMG module is
\begin{equation}
\mathcal{L}_{DMG}={\parallel \mathbf{R}-\mathbf{F}_d\parallel}_2^2
\end{equation}

\subsection{Spatial-Spectral Residual Compensation Module}\label{subsec_SSRCM}

There are various land covers in remote sensed data and the spatial details contained in PAN are abundant. So, directly obtaining $\mathbf{m}^h$ and $\mathbf{n}^h$ through up-sampling of $\mathbf{m}^l$ and $\mathbf{n}^l$ inevitably leads to spatial loss. Therefore, we use an SSRC module to enhance the spatial resolution and at the same time compensate the spectral information in the second step.

\begin{figure}[t]
	\centering
	\includegraphics[width=0.8\textwidth]{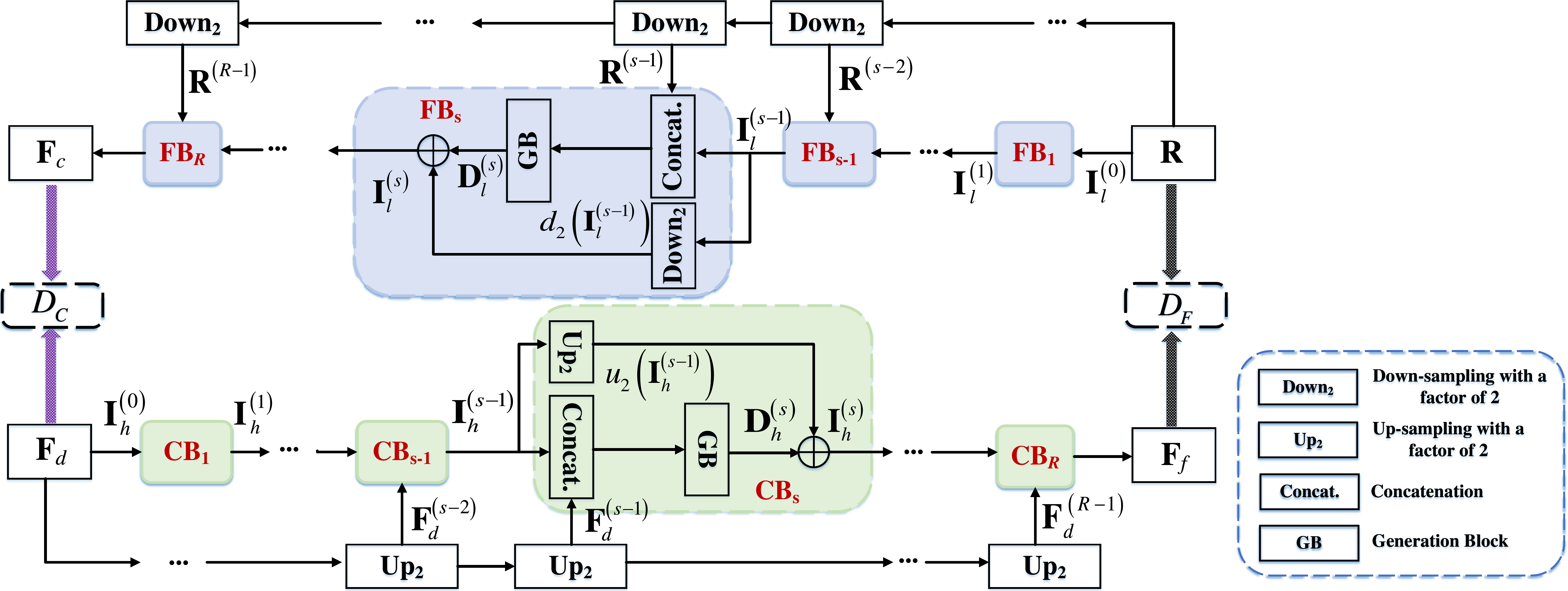}
	\caption{Architecture of SSRC module. It is composed of a series of coarse block (CB) and fine block (FB).}\label{fig4}
\end{figure}

The network architecture for SSRC module is shown in Figure \ref{fig4}. It has several coarse blocks (CBs) and fine blocks (FBs), and each CB has an up-sampling, a concatenate operation and a generator block (GB), while the FB is composed of a down-sampling, a concatenate operation and a GB.

Let $u_2(\cdot)$ and $d_2(\cdot)$ be the up-sampling and down-sampling operations. $u_2(\cdot)$ smooths and expands the $j \times j$ image to be the $2j 
\times 2j$ image, and $d_2(\cdot)$ blurs and decimates the $j \times j$ image to the $j / 2 \times j / 2$ image. For CBs, let $\{\mathbf{F}_d^{(0)},..., \mathbf{F}_d^{(R-1)}\}$, $\{\mathbf{I}_h^{(0)},...,\mathbf{I}_h^{(R)}\}$ and $\{\mathbf{D}_h^{(1)},...,\mathbf{D}_h^{(R)}\}$ denote the input images, the output images and the spatial details produced by GB. We firstly obtain $\{\mathbf{F}_d^{(0)},...,\mathbf{F}_d^{(R-1)}\}$ by progressively up-sampling the fusion results of DMG:
\begin{equation}
\begin{aligned}
&\mathbf{F}_d^{(0)}=\mathbf{F}_d \\
&\mathbf{F}_d^{(s)}=\underbrace{u_2\Big(u_2\big(\ldots u_2(\mathbf{F}_d)\big)\Big)}_s, s=1,...,R-1
\end{aligned}
\end{equation}
where $s$ is the up-sampling level, and $R$ is related to the spatial resolution ratio between images in the coarse domain and the fine domain. Suppose that $\mathbf{I}_h^{(0)}=\mathbf{F}_d$, then, the spatial details are obtained through
\begin{equation}
\mathbf{D}_h^{(s)}=G_s\Big({\Big[\mathbf{I}_h^{(s-1)}, \mathbf{F}_d^{(s-1)}\Big]}_{dim}\Big),
\end{equation}
in which ${[\mathbf{A},\mathbf{B}]}_{dim}$ means the concatenation of $\mathbf{A}$ and $\mathbf{B}$ in the spectral dimension and $G_s(\cdot)$ is the generator at the scale $s$ .

We can obtain $\{\mathbf{I}_h^{(1)},...,\mathbf{I}_h^{(s)},...,\mathbf{I}_h^{(R)}\}$ by
\begin{equation}
\mathbf{I}_h^{(s)}=\frac{u_2(\mathbf{I}_h^{(s-1)})+\mathbf{D}_h^{(s)}}{2}
\end{equation}
\par
After a cascade of CBs, the refined fusion result is
\begin{equation}
\mathbf{F}_f=\mathbf{I}_h^{(R)}.
\end{equation}

Similarly, for FBs, the input images are \{$\mathbf{R}^{(0)},...,\mathbf{R}^{(R-1)}$\}, and we use \{$\mathbf{D}_l^{(1)},...,{\mathbf{D}_l}^{(R)}$\} and \{$\mathbf{I}_l^{(0)},...,\mathbf{I}_l^{(R)}$\} to denote the detail images and output images. The input images are acquired by
\begin{equation}
\begin{aligned}
&\mathbf{R}^{(0)}=\mathbf{R} \\
&\mathbf{R}^{(s)}=\underbrace{d_2\Big(d_2\big(\ldots d_2(\mathbf{R})\big)\Big)}, s=1,...,R-1
\end{aligned}
\end{equation}
and suppose ${\mathbf{I}_l}^{(0)}=\mathbf{R}$, the detail images are
\begin{equation}
\mathbf{D}_l^{(s)}=G_s\big({\big[{I_l}^{(s-1)}, \mathbf{R}^{(s-1)}\big]}_{dim}\big),
\end{equation}
then the output images are
\begin{equation}
I_l^{(s)} = \frac{d_2(\mathbf{I}_l^{(s-1)})+\mathbf{D}_l^{(s)}}{2}.
\end{equation}

Finally, the output of C2F is
\begin{equation}
\mathbf{F}_c={\mathbf{I}_l}^{(R)}.
\end{equation}

\subsection{Joint Compensation Loss Function}\label{subsec_JCLF}

Based on the specific structure of PC-GANs, we propose a joint compensation loss function that enables the triple GANs to be trained simultaneously.

Roughly, we formulate the joint compensation loss function $\mathcal{L}^{JC}$ as the weighted sum of an adversarial loss $\mathcal{L}^{Adv}$ , a cycle-consistent loss $\mathcal{L}^{Cyc}$ and a reconstruction loss $\mathcal{L}^R$
\begin{equation}
\mathcal{L}^{JC}=\mathcal{L}^{Adv}+\lambda_1\mathcal{L}^{Cyc}+\lambda_2\mathcal{L}^R
\end{equation}
where $\lambda_1$ and $\lambda_2$ control the relative importance of the cycle-consistent loss and the reconstruction loss.

For presentation clarity, DMG, C2F and F2C are represented by $G_{DMG}(\cdot)$, $D_{C2F}(\cdot)$ and $G_{F2C}(\cdot)$, respectively. In the following, we detail the loss function for PC-GANs.

\emph{(1) Adversarial Loss: }In this paper, we adopt the least-squares GAN (LSGAN) \cite{mao2017least} objective for training, and the training objectives for DMG, C2F and F2C are
\begin{equation}
\begin{aligned}
&\mathcal{L}_{G_{DMG}}^{Adv}=\frac{1}{2}{\Big(D_C\big(G_{DMG}(\mathbf{I}_1,\mathbf{I}_2)\big)-a\Big)}^2 \\
&\mathcal{L}_{G_{C2F}}^{Adv}=\frac{1}{2}{\bigg(D_F\Big(G_{C2F}\big(G_{DMG}(\mathbf{I}_1,\mathbf{I}_2)\big)\Big)-a\bigg)}^2 \\
&\mathcal{L}_{G_{F2C}}^{Adv}=\frac{1}{2}{\Big(D_C\big(G_{F2C}(\mathbf{R})\big)-a\Big)}^2 \\
\end{aligned}
\end{equation}
where $a$ denotes the value that the generators want the discriminators to believe for fake samples.

For discriminators, the training objectives are
\begin{equation}
\begin{aligned}
&\mathcal{L}_{D_C}^{DMG}=\frac{1}{2}{\Big(D_C\big(G_{DMG}(\mathbf{I}_1,\mathbf{I}_2)\big)-b\Big)}^2 +\frac{1}{2}{\Big(D_C\big(G_{F2C}(\mathbf{R})\big)-c\Big)}^2 \\
&\mathcal{L}_{D_C}^{F2C}=\frac{1}{2}{\Big(D_C\big(G_{DMG}(\mathbf{I}_1,\mathbf{I}_2)\big)-c\Big)}^2 +\frac{1}{2}{\Big(D_C\big(G_{F2C}(\mathbf{R})\big)-b\Big)}^2 \\
&\mathcal{L}_{D_F}=\frac{1}{2}{\bigg(D_F\Big(\mathbf{R}\Big)-c\bigg)}^2 +\frac{1}{2}{\bigg(D_F\Big(G_{C2F}\big(G_{DMG}(\mathbf{I}_1,\mathbf{I}_2)\big)\Big)-b\bigg)}^2,
\end{aligned}
\end{equation}
where $b$ and $c$ are the labels for fake and real samples. We set $a=1$, $b=0$ and $c=1$ in our method according to the advice given in \cite{mao2017least}.

\emph{(2) Reconstruction Loss:} Basically, the generated images should have the most similar spatial structures and spectral information with the images in the same domain. Therefore, the reconstruction loss is introduced
\begin{equation}
\begin{aligned}
&\mathcal{L}_{G_{C2F}}^R=\frac{1}{2}{\parallel G_{C2F}\big(G_{DMG}(\mathbf{I}_1, \mathbf{I}_2)\big)-\mathbf{R}\parallel}_2^2 \\
&\mathcal{L}_{G_{F2C}}^R=\frac{1}{2}{\parallel G_{F2C}(\mathbf{R})-G_{DMG}(\mathbf{I}_1, \mathbf{I}_2)\parallel}_2^2
\end{aligned}
\end{equation}

\emph{ (3) Cycle-Consistent Loss:}Due to the fact that our PC-GANs model has a cyclic structure, it will benefit a lot from the cycle-consistent loss \cite{zhu2017unpaired}. It is supposed that $\mathbf{F}_d=G_{DMG}(\mathbf{I}_1,\mathbf{I}_2)$, and based on the structure of PC-GANs, the transformed images can be brought back to their original forms, i.e., $\mathbf{F}_d\rightarrow G_{C2F}(\mathbf{F}_d)\rightarrow G_{F2C}(G_{C2F}(\mathbf{F}_d))\approx \mathbf{F}_d$.

Then the cycle-consistent loss for PC-GANs is defined as
\begin{equation}
\begin{aligned}
&\mathcal{L}_{G_{C2F}}^{Cyc}=\frac{1}{2}{\parallel G_{F2C}\Big(G_{C2F}\big(G_{DMG}(\mathbf{I}_1, \mathbf{I}_2)\big)\Big)-G_{DMG}(\mathbf{I}_1, \mathbf{I}_2)\parallel}_2^2 \\
&\mathcal{L}_{G_{F2C}}^{Cyc}=\frac{1}{2}{\parallel G_{C2F} \big(G_{F2C}(\mathbf{R})\big)-\mathbf{R}\parallel}_2^2.
\end{aligned}
\end{equation}

In conclusion, PC-GANs can be trained jointly. Owing to the fact that the DMG generator has been pre-trained by the strong supervision, we train the DMG firstly by alternately optimizing $\mathcal{L}_{D_c}^{DMG}$ and $\mathcal{L}_{D_{DMG}}^{Adv}$. After training the DMG, C2F and F2C are optimized by the following joint loss functions
\begin{equation}
\begin{aligned}
&\mathcal{L}_{G_{C2F}}^{JC}=\mathcal{L}_{G_{C2F}}^{Adv}+\lambda_1\mathcal{L}_{G_{C2F}}^{Cyc}+\lambda_2\mathcal{L}_{G_{C2F}}^R \\
&\mathcal{L}_{G_{F2C}}^{JC}=\mathcal{L}_{G_{F2C}}^{Adv}+\lambda_1\mathcal{L}_{G_{F2C}}^{Cyc}+\lambda_2\mathcal{L}_{G_{F2C}}^R.
\end{aligned}
\end{equation}

When training the F2C, $\mathcal{L}_{D_C}^{F2C}$ and $\mathcal{L}_{G_{F2C}}^{JC}$ are optimized, whereas, the C2F is trained by minimizing $\mathcal{L}_{D_F}$ and $\mathcal{L}_{G_{C2F}}^{JC}$ alternately. It should be noted that the DMG is also trained along with the optimization of F2C and C2F. Therefore, the training of PC-GANs is a joint cyclic training strategy.

\section{Experimental Results and Analysis}\label{sec_EXP}

In this section, some experimental results will be presented to demonstrate the effectiveness of our proposed PC-GANs for pan-sharpening.  First, we briefly introduce the datasets. Then, we give the analysis of network configuration and the investigations of some hyper-parameters. Finally, seven state-of-the-art pan-sharpening methods including the Gram–Schmidt adaptive (GSA) \cite{aiazzi2007improving}, generalized Laplacian pyramid with MTF-matched filter and high pass filtering injection model (MTF-GLP-HPM) \cite{aiazzi2003mtf}, two-step sparse coding method with patch normalization (PN-TSSC) \cite{jiang2013two}, target-adaptive PNN (TA-PNN) \cite{scarpa2018target}, DML-GMME \cite{xing2018pan}, Gradient Transformation Prior for Pansharpening (GTP-PNet) \cite{zhang2021gtp} and PanGAN \cite{ma2020pan}, are used for comparisons both in quantitative and visual aspects.  Among them, GSA \cite{aiazzi2007improving} belongs to the CS-based methods, and MTF-GLP-HPM \cite{aiazzi2003mtf} belongs to the MRA-based methods. PN-TSSC \cite{jiang2013two} is the OR-based method. TA-PNN \cite{scarpa2018target}, DML-GMME \cite{xing2018pan}, GTP-PNet \cite{zhang2021gtp} and PanGAN \cite{ma2020pan} are in the family of DL-based approaches.

\begin{figure}[t]
	\centering
	\subfigure[]{\includegraphics[scale=1.3]{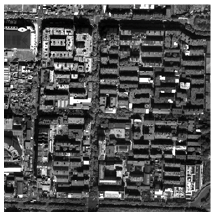}}
	\subfigure[]{\includegraphics[scale=1.3]{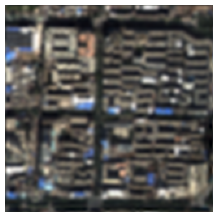}}
	\subfigure[]{\includegraphics[scale=1.3]{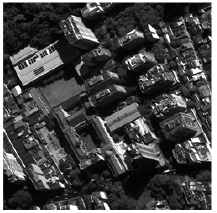}}
	\subfigure[]{\includegraphics[scale=1.3]{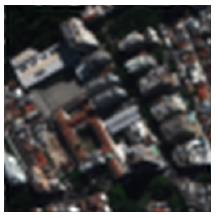}}
	\caption{Examples of test PAN and MS (EXP) images. (a) QuickBird PAN. (b) QuickBird MS (RGB). (c) WorldView-4 PAN. (d) WorldView-4 MS (RGB). }\label{fig5}
\end{figure}

\subsection{Datasets}\label{subsec_DATASET}

Due to that the characteristics of RS images from different satellites are different, we train our network on two datasets independently, i.e., QuickBird (QB) and WorldView-4 (WV-4). For each dataset, there are totally 48900 patches that are randomly sampled from the corresponding satellite datasets, and we choose 39120 patches for training while the rest 9780 patches for validation. Afterwards, we crop another 200 test image pairs that are not contained in the training and validation set to form a test set. The test MS images have the size of $256 \times 256 \times 4$  and PAN images are composed of $1024 \times 1024$ pixels.

Figure \ref{fig5} shows some test data illustrated in our experiments, which are collected from QB and WV-4 satellites, and the red (R), green (G) and blue (B) bands are displayed for visualization. The MS images shown in Fig. 5 are interpolated by a polynomial kernel with 23 coefficients (EXP) \cite{vivone2014critical}. Figure \ref{fig5}(a) and Figure \ref{fig5}(b) represent the urban area of Xi’an (China) from QB satellite, and the resolution is 2.44m and 0.61m for MS and PAN image, respectively. Figure \ref{fig5}(c) and Figure \ref{fig5}(d) show areas of Rio de Janeiro (Brazil) captured by WV-4 satellite \footnote{Available online: https://www.digitalglobe.com}  in May 2016, where the resolution of PAN and MS is 0.31m and 1.24m respectively. In the reduced resolution case, the original MS and PAN images are filtered by a Gaussian filter that is matched to the modulation transfer function (MTF) of sensors, and then they are down-sampled with a decimation factor of 4. After that, the original MS images are used as reference images according to Wald’s protocol \cite{wald1997fusion}. The availability of reference images allows for the use of several widespread assessment indexes, such as Q4, SAM and ERGAS \cite{vivone2014critical}. In the full resolution case, original MS and PAN are directly taken as inputs without other down-sampling or filtering operations. Therefore, there are no reference images, so we evaluate the fusion results by  $D_\lambda$, $D_s$   and QNR indexes \cite{alparone2008multispectral}. The ideal value of SAM, ERGAS,  $D_\lambda$ and $D_s$  is zero, while that of Q4 and QNR is one.

\subsection{Experimental Settings}\label{subsec_EXPSET}

\emph{1) Training Details:} In order to train the model, reference images are required, therefore, we filter the original PAN and MS images with the MTF-matched Gaussian filter and down- sample them with a decimation factor of 4, then the original MS images are treated as reference images and the down-sampled MS and PAN are taken as the inputs for PC-GANs.

During training, the patch size is   and the batch size is set to 16. The weights are initialized from a Gaussian distribution $N(0, 0.02)$. We use the Adam optimizer for generators and mini-batch gradient descent optimizer for discriminators to optimize the network parameters, where $\beta_1=0.5$ and $\beta_2=0.999$ in Adam algorithm. The learning rate is initialized as 0.0002 and it is divided by 2 every 100 epochs. The implementation is supported by the Tensorflow framework and the training is conducted on two GPUs (Nvidia GTX 1080Ti). The model has been trained 300 epochs in total, and the training time is about four hours.

\begin{table}[ht]
	\centering
	\caption{Configuration of DMG Module}\label{table_1}
	
	\begin{tabular}{cccccc}
		\hline
		Network                          & Layer & Kernel Size        & Dilation & Output Channels & Nonlinearity            \\ 
		\hline
		\multirow{8}{*}{$f(\cdot)$} & 1     & $3\times3$         & 1        & 32              & LReLU                   \\
		& 2     & $3\times3$         & 1        & 32              & LReLU                   \\
		& 3     & $3\times3$         & 2        & 32              & LReLU                   \\
		& 4     & $3\times3$         & 4        & 32              & LReLU                   \\
		& 5     & $3\times3$         & 8        & 32              & LReLU                   \\
		& 6     & $3\times3$         & 16       & 32              & LReLU                   \\
		& 7     & $3\times3$         & 1        & 32              & LReLU                   \\
		& 8     & $1\times1$         & 1        & 4               & -                 \\ \cline{1-6} 
	\end{tabular}
\end{table}

\begin{figure}[ht]
	\centering
	\subfigure[]{\includegraphics[scale=0.71]{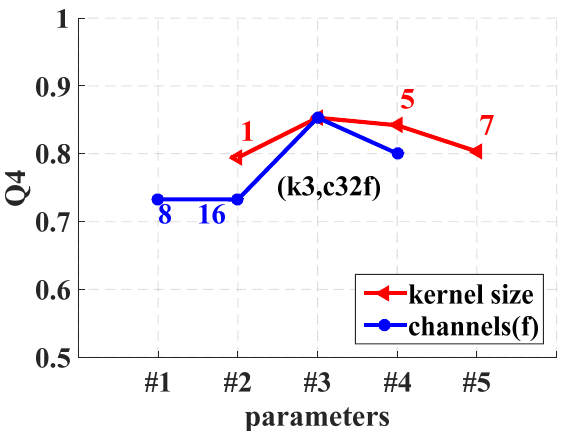}}
	\subfigure[]{\includegraphics[scale=0.71]{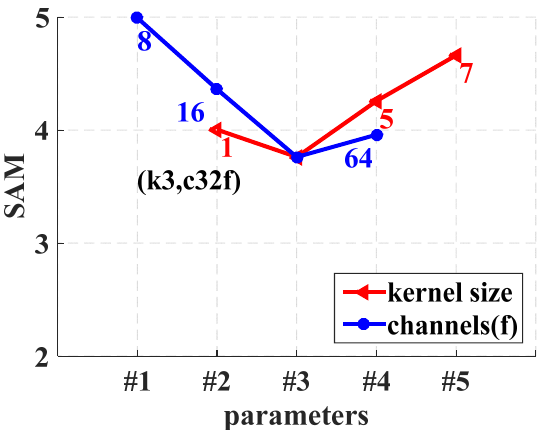}}
	\subfigure[]{\includegraphics[scale=0.71]{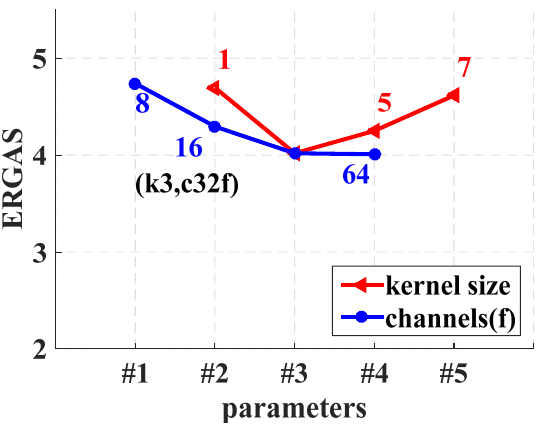}}
	\caption{Different parameter configurations of DMG module with respect to (a) Q4 index, (b) SAM index, and (c) ERGAS index.}\label{fig6}
\end{figure}

\begin{table}[h]
	\centering
	\caption{\protect\centering{Configuration of SSRC Module}}\label{table_2}
	\resizebox{\textwidth}{50mm}{
		\begin{tabular}{c|c|c|c|c|c|c}
			\hline
			\hline
			Network                      & \multicolumn{2}{c|}{Layer}                     & Kernel Size        & Stride  & Output Channels & Nonlinearity \\ \hline \hline
			\multirow{8}{*}{GB}          & \multicolumn{2}{c|}{C1}                        & $3\times3$         & 2       & 4               & ReLU         \\ \cline{2-7}
			& \multicolumn{2}{c|}{C2}                        & $3\times3$         & 2       & 8               & ReLU         \\ \cline{2-7}
			& \multicolumn{2}{c|}{C3}                        & $3\times3$         & 2       & 16              & ReLU         \\ \cline{2-7}
			& $RB_x$                           & C1  & $3\times3$         & 1       & 16              & ReLU         \\ \cline{3-7}
			&  $x=1,...,N$                     & C2  & $3\times3$         & 1       & 16              & -      \\  \cline{2-7}
			& \multicolumn{2}{c|}{D1}                        & $3\times3$         & 1/2     & 16              & ReLU         \\  \cline{2-7}
			& \multicolumn{2}{c|}{D2}                        & $3\times3$         & 1/2     & 8               & ReLU         \\  \cline{2-7}
			& \multicolumn{2}{c|}{D3}                        & $3\times3$         & 1/2     & 4               & -      \\ \hline \hline
			\multirow{7}{*}{$D_C/D_F$}  & \multicolumn{2}{c|}{C1}                        & $3\times3$         & 1       & 32              & LReLU        \\ \cline{2-7}
			& \multicolumn{2}{c|}{C2}                        & $3\times3$         & 2       & 64              & LReLU        \\ \cline{2-7}
			& \multicolumn{2}{c|}{C3}                        & $3\times3$         & 2       & 128             & LReLU        \\ \cline{2-7}
			& \multicolumn{2}{c|}{C4}                        & $3\times3$         & 2       & 256             & LReLU        \\ \cline{2-7}
			& \multicolumn{2}{c|}{C5}                        & $3\times3$         & 2       & 512             & LReLU        \\ \cline{2-7}
			& \multicolumn{2}{c|}{C6}                        & $3\times3$         & 1       & 512             & LReLU        \\ \cline{2-7}
			& \multicolumn{2}{c|}{C7}                        & $3\times3$         & 1       & 1               & -      \\
			\hline
			\hline      
	\end{tabular}}
\end{table}

\begin{figure}[t]
	\centering
	\subfigure{\includegraphics[scale=0.7]{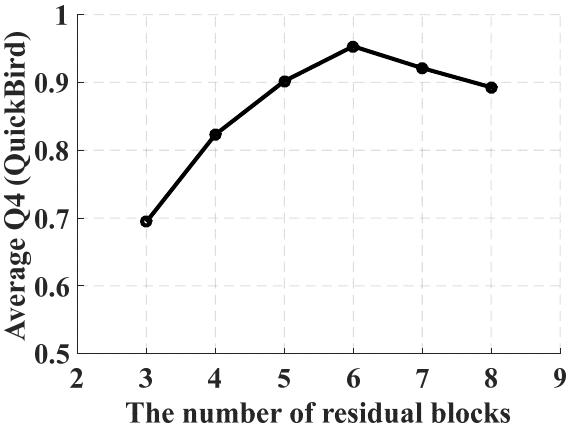}}
	\subfigure{\includegraphics[scale=0.7]{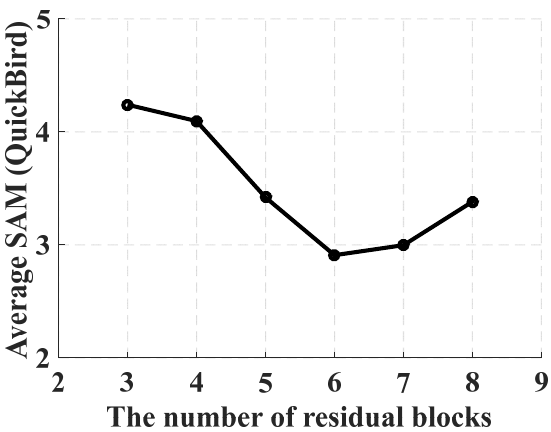}}
	\subfigure{\includegraphics[scale=0.7]{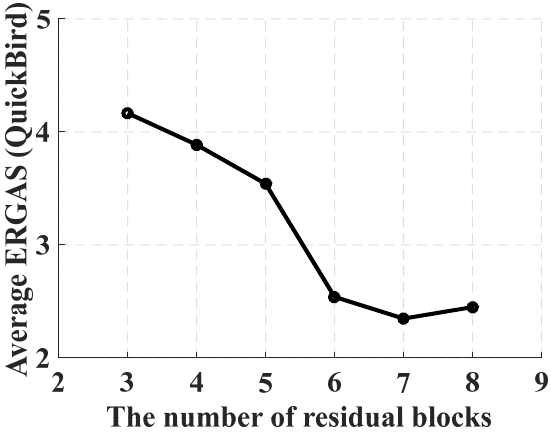}}
	\caption{Quality indexes with respect to the different number of residual blocks.}\label{fig7}
\end{figure}

\emph{2) Network Architectures:} The configuration of the DMG module is shown in Table~\ref{table_1}, where the dilation rates of the first two layers together with the last two layers are all set to 1, while they are increased from 2 to 16 for the middle four layers. When the dilation rate equals 16, the receptive field of layer six is $65 \times 65$ and it finally increases to $67 \times 67$ due to a followed $3 \times 3$ convolution, which is enough for our $64 \times 64$ training samples. It should be noted that the training patches are padded by the reflection padding, that is, the buffer zone is filled by reflecting the image about each edge. Moreover, the effects of channel numbers and kernel size of $f(\cdot)$  are also explored. Figure \ref{fig6} shows the average evaluation indexes on the QB dataset with different network configurations. For a specific index, one of the parameters is adjusted with another parameter fixed and a changing curve is obtained. For clarity, we shift the blue curve horizontally to make them intersect at a joint point (k3, c32f), where k3 and c32f mean that the kernel size and channel numbers of $f(\cdot)$ are $3 \times 3$ and 32 respectively. It can be observed from Figure \ref{fig6} that this configuration obtains a tradeoff among Q4, SAM and ERGAS indexes. Moreover, the configuration of SSRC module is shown in Table~\ref{table_2}, where GB and RBx represent the generator block and the x-th residual block within the specific generator block. The letter C (D) in the Layer column means that current layer is a convolutional (deconvolutional) layer. The nonlinear function used in our model is a rectified linear unit (ReLU) or a leaky ReLU (LReLU) with $\alpha=0.2$. From the definition of the Laplacian pyramid, we know that we need $R=log_2r$ CBs to map low-resolution MS images $\mathbf{M}_l \in R^{m \times m \times 4}$ to corresponding high-resolution MS images $\mathbf{M}_h \in R^{M \times M \times 4}$ ($M=r \times m$, and $r=4$ in this paper), so $R=2$ is chosen. In order to guarantee the success of the adversarial training, the discriminators should have enough capacity, therefore, the channel numbers of discriminators double from 32 to 512 with the increase of the network depth for the first six layers, which proved to be effective from several experiments we have taken, and if the channel number of the first layer is 64, the number of parameters to be learned becomes too large while the capacity of the discriminator is not enough if the basic channel number is set to be 16. Therefore, the basic channel number is chosen as 32. It can be observed from Table~\ref{table_2} that there are three convolutional layers before N residual blocks together with other corresponding three deconvolutional layers after the residual blocks. Such convolutional or deconvolutional layers act as down-sampling or up-sampling operations. We have conducted experiments on the QB dataset to investigate the influence of the number of residual blocks. The investigation is shown in Figure \ref{fig7}, where Q4 and SAM obtain the best mean values when $N=6$ while ERGAS obtains its best value when  $N=7$  Finally, we choose $N=6$ , because Q4 and SAM results are the best and the ERGAS result is also admissible in this configuration.

\begin{figure}[ht]
	\centering
	\subfigure[]{\includegraphics[scale=0.74]{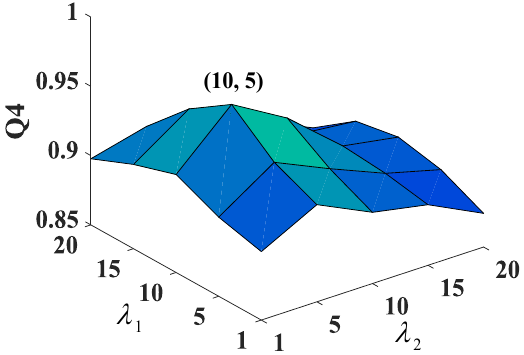}}
	\subfigure[]{\includegraphics[scale=0.74]{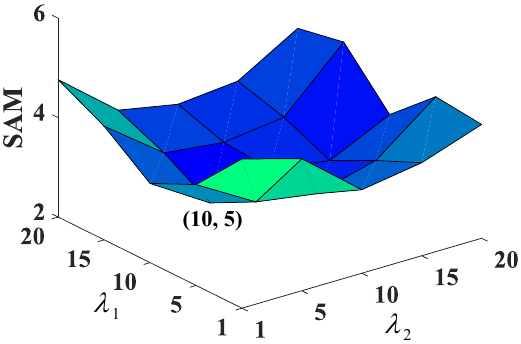}}
	\subfigure[]{\includegraphics[scale=0.74]{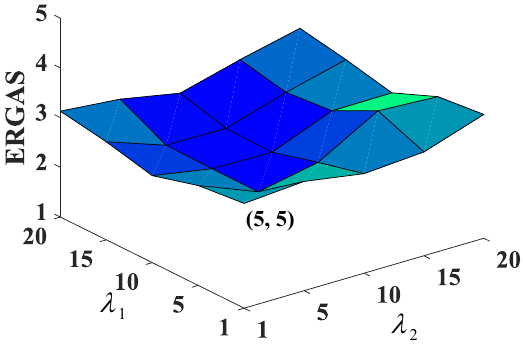}}
	\caption{Average quality index with different values of   and  . (a) Variations on Q4. (b) Variations on SAM. (c) Variations on ERGAS.}\label{fig8}
\end{figure}

Finally, experiments are also conducted to investigate the influence of $\lambda_1$ and $\lambda_2$ which control the importance of cycle loss and reconstruction loss during training. Related studies are shown Figure \ref{fig8}. For Q4 and SAM, they obtain the extreme value at the position (10, 5) which means that $\lambda_1=10$ and  $\lambda_2=5$, while from Figure \ref{fig8}(c), we can observe that ERGAS obtains the extreme value at (5, 5).  Although the extreme point of ERGAS is (5, 5), we still choose $\lambda_1=10$ and  $\lambda_2=5$ in our training, because the variations of $\lambda_1$ and $\lambda_2$ seem to have less influence on ERGAS compared with SAM and Q4.

\begin{figure}[t]
	\centering
	\subfigure[]{\includegraphics[width=0.17\textwidth]{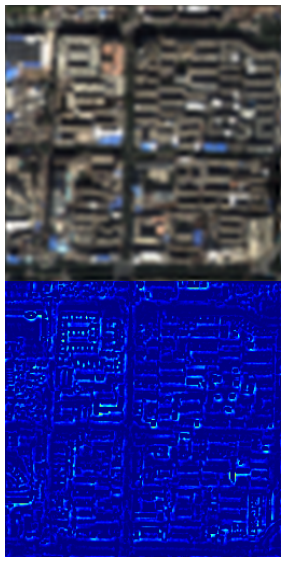}}
	\subfigure[]{\includegraphics[width=0.17\textwidth]{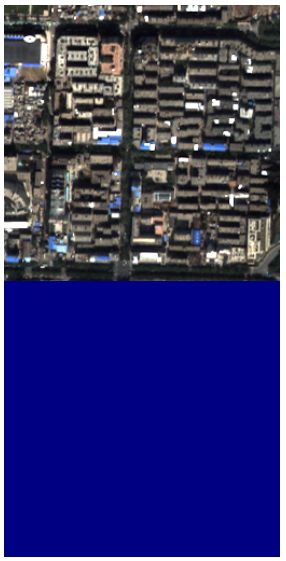}}
	\subfigure[]{\includegraphics[width=0.17\textwidth]{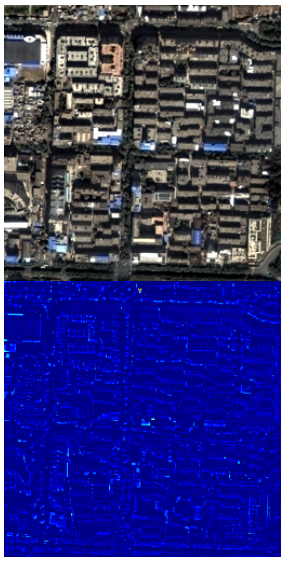}}
	\subfigure[]{\includegraphics[width=0.17\textwidth]{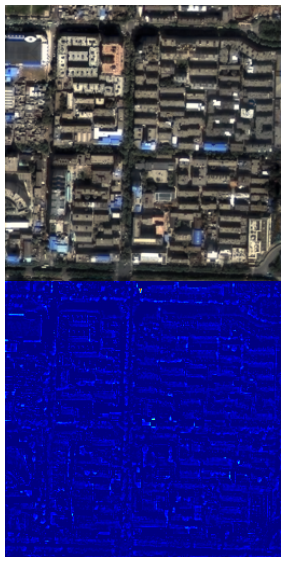}}
	\subfigure[]{\includegraphics[width=0.17\textwidth]{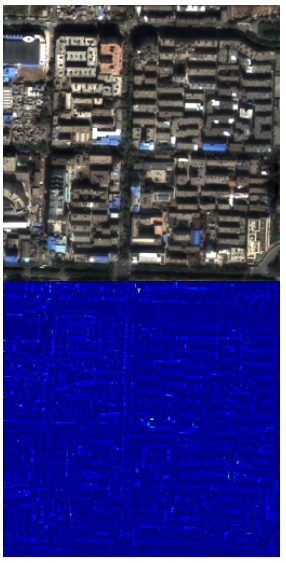}}
	\subfigure[]{\includegraphics[width=0.17\textwidth]{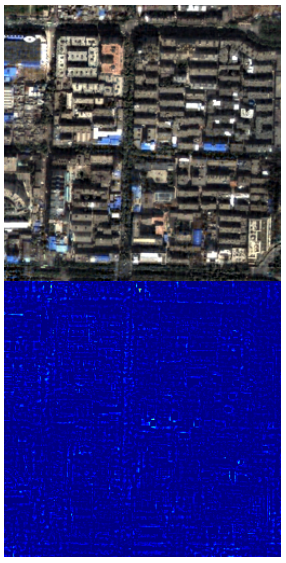}}
	\subfigure[]{\includegraphics[width=0.17\textwidth]{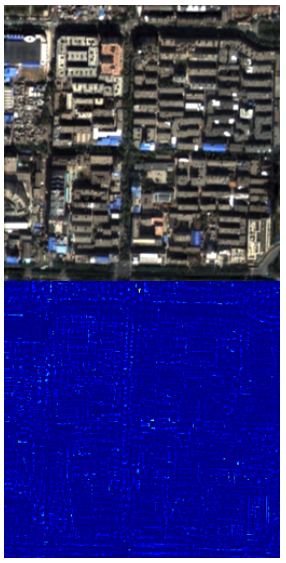}}
	\subfigure[]{\includegraphics[width=0.17\textwidth]{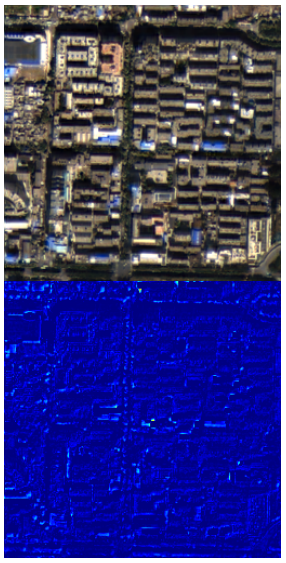}}
	\subfigure[]{\includegraphics[width=0.17\textwidth]{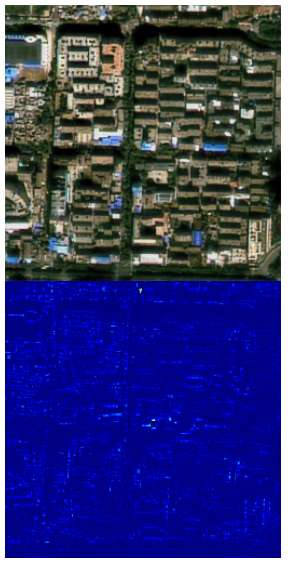}}
	\subfigure[]{\includegraphics[width=0.17\textwidth]{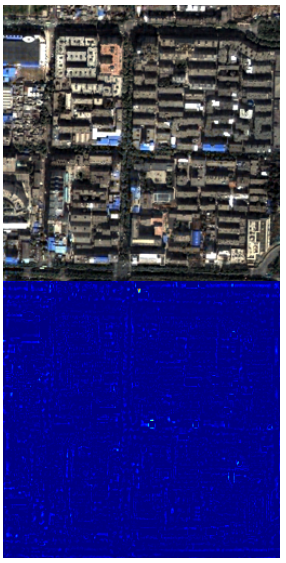}}
	\subfigure{\includegraphics[width=0.5\textwidth]{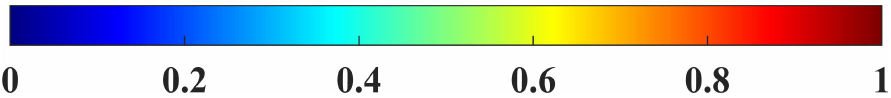}}
	\caption{Fusion results and error maps on QuickBird data at the reduced resolution. (a) Original MS (up-sampled). (b) Reference. The fusion results of (c) GSA \cite{aiazzi2007improving}, (d) MTF-GLP-HPM \cite{aiazzi2003mtf}, (e) PN-TSSC \cite{jiang2013two}, (f) TA-PNN \cite{scarpa2018target}, (g) DML-GMME \cite{xing2018pan}, (h) GTP-PNet \cite{zhang2021gtp}, (i) PanGAN \cite{ma2020pan}, and (j) PC-GANs, where the first row shows the original images or the fusion results and the second row of (a)-(j) shows the error maps between reference image and fusion results.}\label{fig9}
\end{figure}

\subsection{Experimental Results at Reduced Resolution}\label{subsec_ERRR}

\begin{table}[h]
	\centering
	\caption{\protect\centering{Evaluation Indexes of Different Methods on QuickBird Data at Reduced Resolution}}\label{table_3}
	\setlength{\tabcolsep}{7mm}{
		\begin{tabular}{cccc}
			\hline
			\hline
			Method      & Q4     & SAM    & ERGAS  \\ \hline \hline
			GSA         & 0.9456 & 2.9614 & 3.0288 \\ \hline
			MTF-GLP-HPM & 0.9479 & 3.2440 & 2.9126 \\ \hline
			PN-TSSC     & 0.9526 & 3.8595 & 3.0661 \\ \hline
			TA-PNN      & 0.9564 & 3.1055 & 2.6500 \\ \hline
			DML-GMME    & 0.9605 & 2.9060 & 2.7298 \\ \hline
			GTP-PNet    & 0.9529 & 3.8366 & 2.8074 \\ \hline
			PanGAN      & 0.9427 & 4.5248 & 2.9539 \\ \hline
			PC-GANs     & $\bm{0.9773}$ & $\bm{2.8485}$ & $\bm{2.5370}$ \\ \hline \hline
	\end{tabular}}
\end{table}

In this section, our PC-GANs based pan-sharpening method is compared with several state-of-the-art pan-sharpening methods on datasets acquired from QB, and WV-4 satellites at the reduced resolution. Figure \ref{fig9} and Table~\ref{table_3} show the fusion results of comparison methods and our proposed method on QB data. Due to the fact that the results are all zoomed out in this paper, we draw the error maps to clearly demonstrate the differences. The error values are first computed in each spectral band and then summed together to achieve the resulting error maps. It is obvious that the result of OR-based method, PN-TSSC \cite{jiang2013two}, and DL-based methods, TA-PNN \cite{scarpa2018target}, DML-GMME \cite{xing2018pan}, GTP-PNet \cite{zhang2021gtp}, PanGAN \cite{ma2020pan} and PC-GANs, are better than traditional CS- and MRA-based methods. One can zoom in the figures to watch the differences, where the results of GSA \cite{aiazzi2007improving}, GTP-PNet \cite{zhang2021gtp} and PanGAN \cite{ma2020pan}  have confronted spectral distortion. TA-PNN \cite{scarpa2018target} and GTP-PNet \cite{zhang2021gtp} have over-enhanced the images, as a result, several artifacts appear in their results. On the contrary, MTF-GLP-HPM \cite{aiazzi2003mtf} and DML-GMME \cite{xing2018pan} results are blurry and loss some spatial details. In Figure \ref{fig9}(j), the spectral information is well preserved and the spatial details are fully contained. Thus, our proposed method has the best fusion product. The same conclusion can be drawn from Table~\ref{table_3} that our method enhances the spatial details and preserves the spectral information at the same time.

\begin{table}[ht]
	\centering
	\caption{\protect\centering{Evaluation Indexes of Different Methods on WorldView-4 Data at Reduced Resolution}}\label{table_4}
	\setlength{\tabcolsep}{7mm}{
		\begin{tabular}{cccc}
			\hline \hline
			Method      & Q4     & SAM    & ERGAS  \\ \hline \hline
			GSA         & 0.8220 & 6.7066 & 6.3613 \\ \hline
			MTF-GLP-HPM & 0.8980 & 6.2763 & 5.8909 \\ \hline
			PN-TSSC     & 0.9009 & 5.9762 & 5.7953 \\ \hline
			TA-PNN      & 0.9172 & 5.7640 & 5.3897 \\ \hline
			DML-GMME    & 0.9024 & 6.2230 & 5.7176 \\ \hline
			GTP-PNet    & 0.9034 & 5.8316 & 5.1840 \\ \hline
			PanGAN      & 0.9068 & 6.6713 & 5.3160 \\ \hline
			PC-GANs     & $\bm{0.9361}$ & $\bm{5.4779}$ & $\bm{4.8823}$ \\ \hline \hline
	\end{tabular}}
\end{table}

\begin{figure}[h]
	\centering
	\subfigure[]{\includegraphics[width=0.17\textwidth]{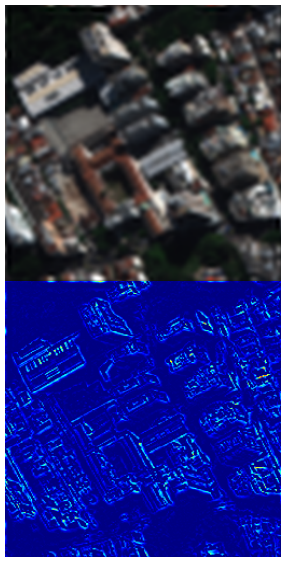}}
	\subfigure[]{\includegraphics[width=0.17\textwidth]{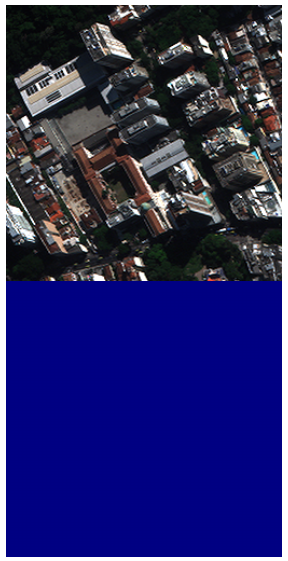}}
	\subfigure[]{\includegraphics[width=0.17\textwidth]{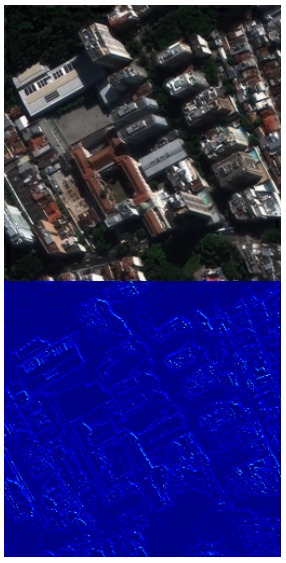}}
	\subfigure[]{\includegraphics[width=0.17\textwidth]{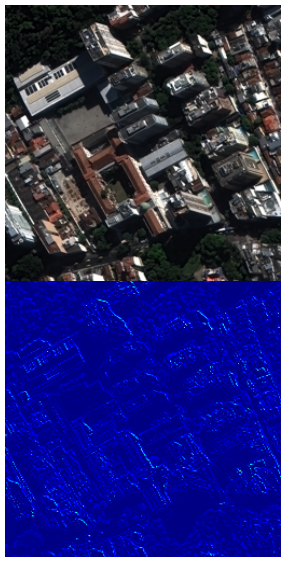}}
	\subfigure[]{\includegraphics[width=0.17\textwidth]{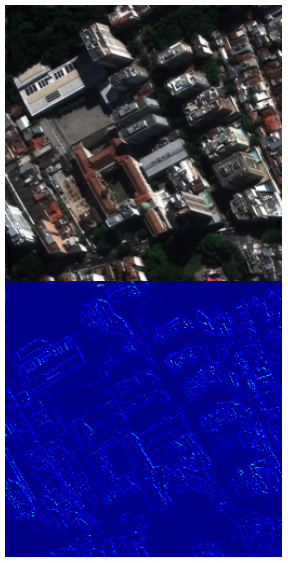}}
	\subfigure[]{\includegraphics[width=0.17\textwidth]{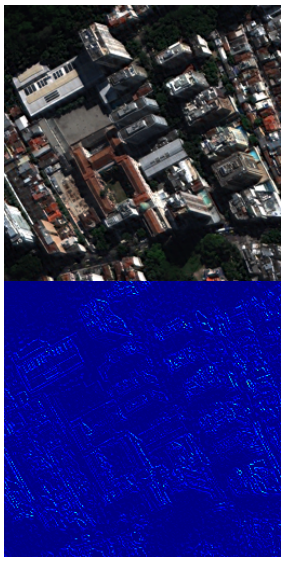}}
	\subfigure[]{\includegraphics[width=0.17\textwidth]{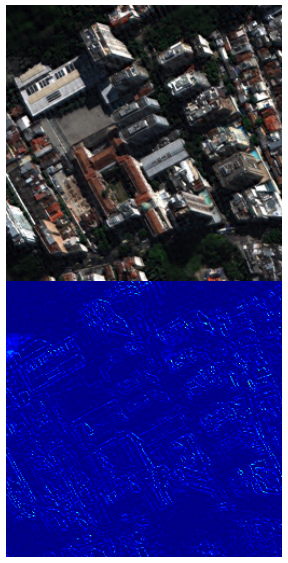}}
	\subfigure[]{\includegraphics[width=0.17\textwidth]{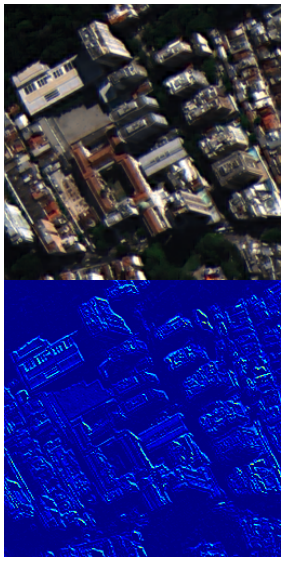}}
	\subfigure[]{\includegraphics[width=0.17\textwidth]{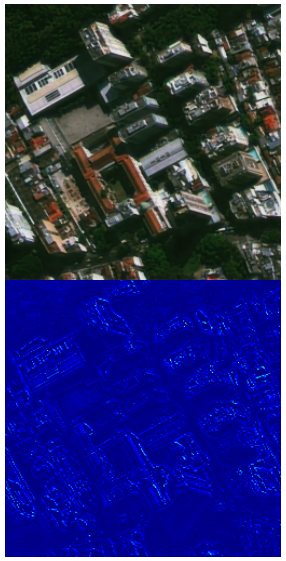}}
	\subfigure[]{\includegraphics[width=0.17\textwidth]{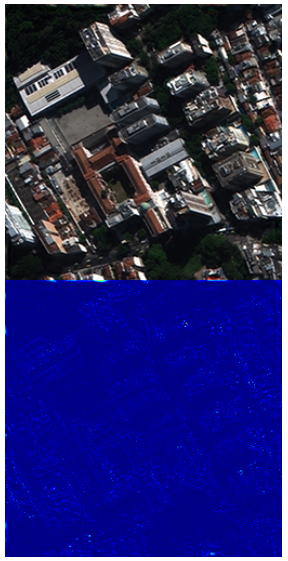}}
	\subfigure{\includegraphics[width=0.5\textwidth]{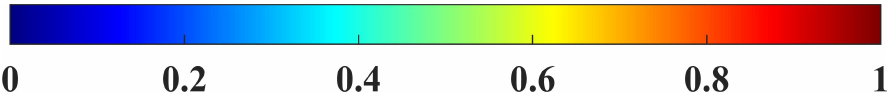}}
	\caption{Fusion results and error maps on WorldView-4 data at the reduced resolution.  (a) Original MS (up-sampled). (b) Reference. The fusion results of (c) GSA \cite{aiazzi2007improving}, (d) MTF-GLP-HPM \cite{aiazzi2003mtf}, (e) PN-TSSC \cite{jiang2013two}, (f) TA-PNN \cite{scarpa2018target}, (g) DML-GMME \cite{xing2018pan}, (h) GTP-PNet \cite{zhang2021gtp}, (i) PanGAN \cite{ma2020pan}, and (j) PC-GANs, where the first row shows the original images or the fusion results and the second row of (a)-(j) shows the error maps between reference image and fusion results.}\label{fig10}
\end{figure}

Moreover, the experiments on WV-4 data are also conducted. The fusion results and quality indexes are shown in Figure \ref{fig10} and Table~\ref{table_4}. Similarly, GSA \cite{aiazzi2007improving}, MTF-GLP-HPM \cite{aiazzi2003mtf}, GTP-PNet \cite{zhang2021gtp}, and PanGAN \cite{ma2020pan} are spectrally distorted. At the same time, the blurry effects and the aliasing effects appear respectively in the results of GSA\cite{aiazzi2007improving}  and MTF-GLP-HPM \cite{aiazzi2003mtf}. Because WV-4 data are high-resolution images, they have more spatial details. The results of PN-TSSC \cite{jiang2013two}, DML-GMME \cite{xing2018pan}, and TA-PNN \cite{scarpa2018target} are slightly blurred, which are also concluded from the error maps that the spatial differences are larger than PC-GANs. The index evaluations listed in Table~\ref{table_4} also in accordance with our visual inspections.

\subsection{Experimental Results at Full Resolution}\label{subsec_ERFR}

In the full resolution experiments, the original MS and PAN are not down-sampled. Due to the fact that there are no reference images for computing the difference images, a certain area of the fused image is zoomed in to verify the effectiveness of our method in the full resolution case.

\begin{figure}[h]
	\centering
	\subfigure[]{\includegraphics[width=0.19\textwidth]{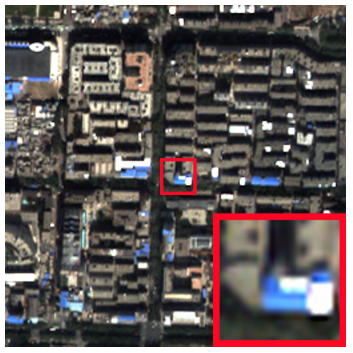}}
	\subfigure[]{\includegraphics[width=0.19\textwidth]{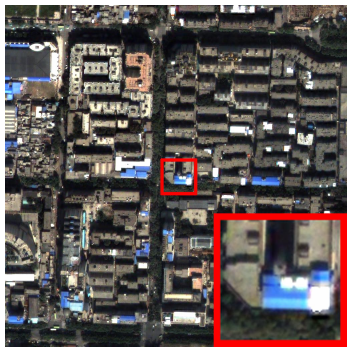}}
	\subfigure[]{\includegraphics[width=0.19\textwidth]{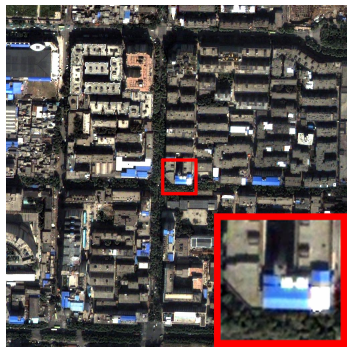}}
	\subfigure[]{\includegraphics[width=0.19\textwidth]{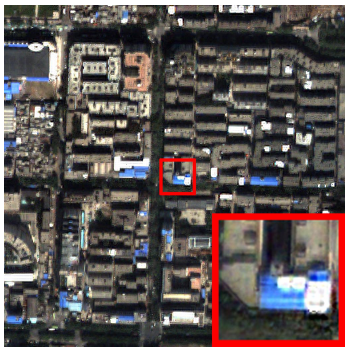}}
	\subfigure[]{\includegraphics[width=0.19\textwidth]{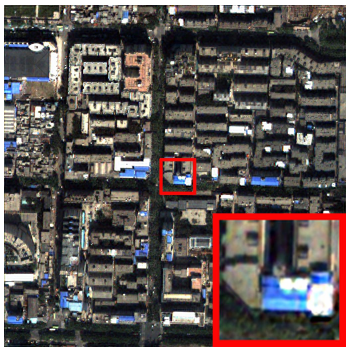}}
	\subfigure[]{\includegraphics[width=0.19\textwidth]{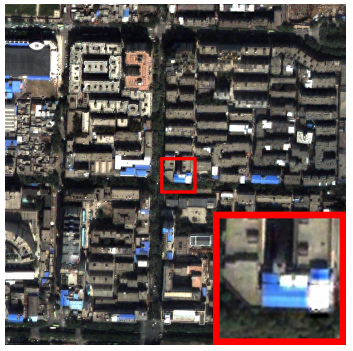}}
	\subfigure[]{\includegraphics[width=0.19\textwidth]{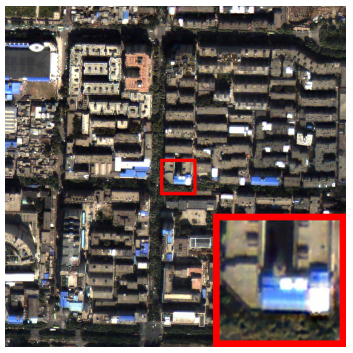}}
	\subfigure[]{\includegraphics[width=0.19\textwidth]{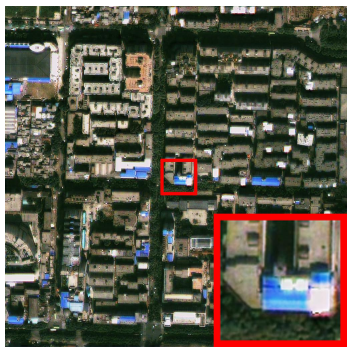}}
	\subfigure[]{\includegraphics[width=0.19\textwidth]{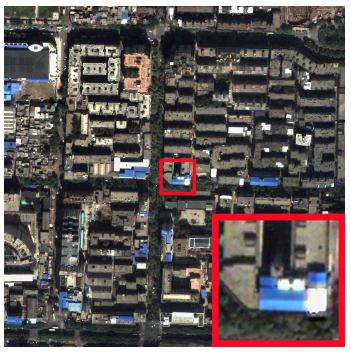}}
	\caption{Fusion results on QuickBird data at full resolution. (a) EXP \cite{vivone2014critical}. (b) GSA \cite{aiazzi2007improving}. (c) MTF-GLP-HPM \cite{aiazzi2003mtf}. (d) PN-TSSC \cite{jiang2013two}. (e) TA-PNN  \cite{scarpa2018target} (f) DML-GMME \cite{xing2018pan}. (g) GTP-PNet \cite{zhang2021gtp}. (h) PanGAN \cite{ma2020pan} (i) PC-GANs.}\label{fig11}
\end{figure}

\begin{figure}[h]
	\centering
	\subfigure[]{\includegraphics[width=0.19\textwidth]{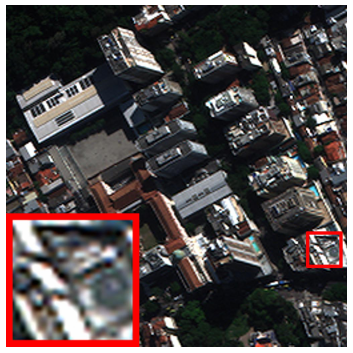}}
	\subfigure[]{\includegraphics[width=0.19\textwidth]{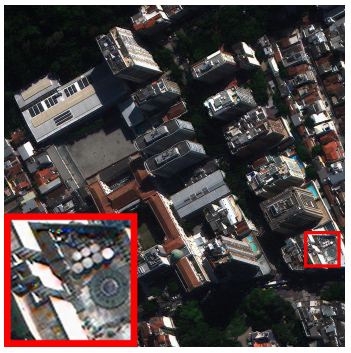}}
	\subfigure[]{\includegraphics[width=0.19\textwidth]{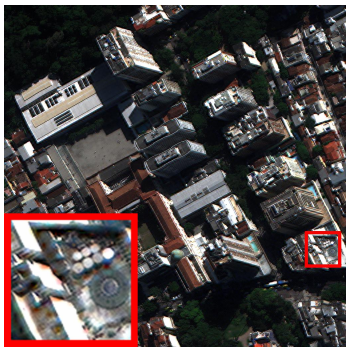}}
	\subfigure[]{\includegraphics[width=0.19\textwidth]{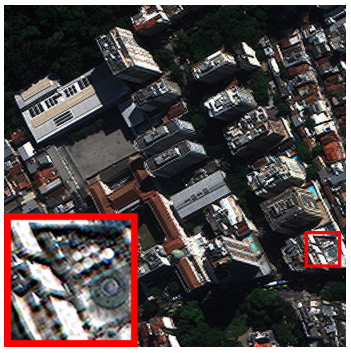}}
	\subfigure[]{\includegraphics[width=0.19\textwidth]{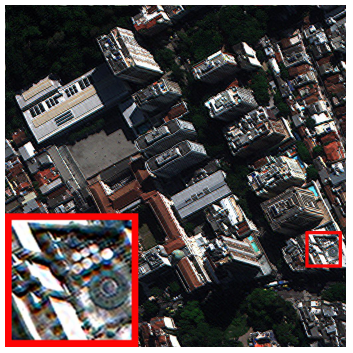}}
	\subfigure[]{\includegraphics[width=0.19\textwidth]{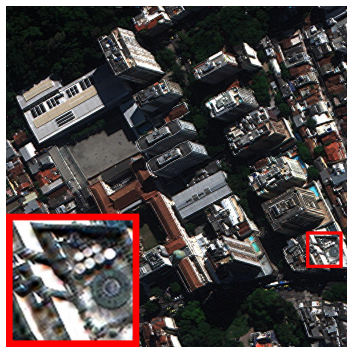}}
	\subfigure[]{\includegraphics[width=0.19\textwidth]{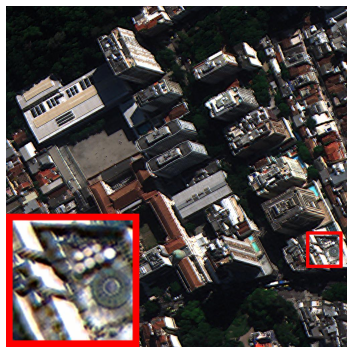}}
	\subfigure[]{\includegraphics[width=0.19\textwidth]{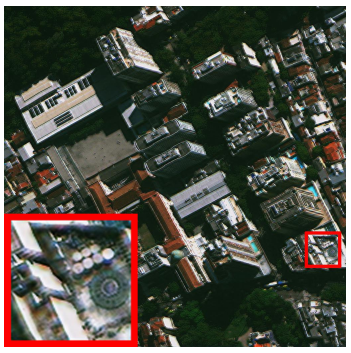}}
	\subfigure[]{\includegraphics[width=0.19\textwidth]{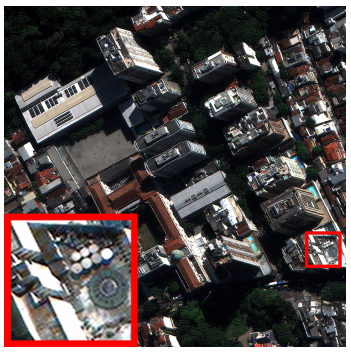}}
	\caption{ Fusion results onWorldView-4 data at full resolution. (a) EXP \cite{vivone2014critical}. (b) GSA \cite{aiazzi2007improving}. (c) MTF-GLP-HPM \cite{aiazzi2003mtf}. (d) PN-TSSC \cite{jiang2013two}. (e) TA-PNN  \cite{scarpa2018target} (f) DML-GMME \cite{xing2018pan}. (g) GTP-PNet \cite{zhang2021gtp}. (h) PanGAN \cite{ma2020pan} (i) PC-GANs.}\label{fig12}
\end{figure}

Figs. \ref{fig11}-\ref{fig12} show the fusion results on QB and WV-4 datasets respectively. For QB dataset, there are some artifacts in Figs. \ref{fig11}(b)-(e), especially in the results of PN-TSSC \cite{jiang2013two} and TA-PNN \cite{scarpa2018target}, while the results of GTP-PNet \cite{zhang2021gtp} and PanGAN \cite{ma2020pan} are seriously spectral distorted. Although the image shown in Fig. \ref{fig11}(f) are comparable to that of our proposed method, it can carefully be seen from the zoomed-in areas that the edges of white areas seem to be a little blurry in Fig. \ref{fig11}(f).

\begin{table}[ht]
	\centering
	\caption{\protect\centering{Evaluation Indexes of Different Methods at Full Resolution}}\label{table_5}
	\begin{tabular}{c|c|c|c|c|c|c}
		\hline \hline
		\multirow{2}{*}{Method} &\multicolumn{2}{c|}{$D_\lambda$} &\multicolumn{2}{c|}{$D_s$} &\multicolumn{2}{c}{QNR}    \\ \cline{2-7}
		&QB        &WV-4                 &QB              &WV-4     &QB               &WV-4     \\ \hline
		GSA                     &0.0306    &0.0817               &0.0302          &\underline{0.0268}   &0.9401           &0.8936   \\ \hline
		MTF-GLP-HPM             &0.0172    &0.0320               &0.0427          &0.0632   &0.9408           &0.9068   \\ \hline
		PN-TSSC                 &0.0325    &0.0227				 &0.0223		  &0.0482   &0.9459           &0.9302   \\ \hline
		TA-PNN                  &0.0192    &0.0315               &0.0285          &0.0556   &0.9528           &0.9146   \\ \hline
		DML-GMME                &0.0197    &\underline{0.0187}               &$\bm{0.0179}$          &0.0308   &\underline{0.9627}           &\underline{0.9511}   \\ \hline
		GTP-PNet                &0.0449    &0.0362               &0.0377          &0.0519   &0.9191           &0.9138   \\ \hline
		PanGAN                  &\underline{0.0143}    &0.0243               &0.0264          &0.0621   &0.9597           &0.9151   \\ \hline
		PC-GANs 				&$\bm{0.0137}$    &$\bm{0.0148}$               &\underline{0.0186}          &$\bm{0.0236}$   &$\bm{0.9680}$           &$\bm{0.9619}$   \\ 
		\hline \hline
	\end{tabular}
\end{table}

For WV-4 data used in this paper, the fusion product GSA \cite{aiazzi2007improving} method has weird color, and spectral distortion appears in the result of PanGAN \cite{ma2020pan}. It can be observed from Figure \ref{fig12}(d)-(f) that the aliasing effects together with the artifacts exist in the results of PN-TSSC \cite{jiang2013two}, TA-PNN \cite{scarpa2018target} and DML-GMME \cite{xing2018pan}. Oppositely, blurry effects arise in the results of MTF-GLP-HPM \cite{aiazzi2003mtf} and GTP-PNet \cite{zhang2021gtp} methods. Finally, our method outperforms other comparison methods and obtains a balance between spatial enhancement and spectral preservation. Quality evaluations given in Table~\ref{table_5} verify our conclusions as well.

\section{Ablation Studies}\label{sec_ABL}

In this section, we conduct a comprehensive ablation study on QB dataset to evaluate the effects of DMG and SSRC modules on the final fusion results. We take the model shown in Figure \ref{fig2} as our Baseline. The ablation study contains the following cases.

\emph{Case 1 (Without DMG module):} In this case, the DMG module is replaced by an average operation, where the MS image is up-sampled and the PAN image is duplicated to make them have the same size, then the average operation is conducted on the MS and PAN image to obtain initial fusion result $\mathbf{F}_d$.

\emph{Case 2 (Without SSRC module):} In this case, we simply abandon the SSRC module, and the model is trained only by the DMG loss $L_{DMG}$. We evaluate the initial fusion result $\mathbf{F}_d$.

\emph{Case 3 (Without C2F generator):} SSRC module is composed of two GANs, and in this case, the C2F generator and the discriminator $D_F$ are discarded. Then the model is bidirectional, where MS and PAN images are fused by the DMG module, and the reference image is used to generate $\mathbf{F}_c$. The model in this case is trained by DMG loss $L_{DMG}$, he reconstruction loss and adversarial loss related only to F2C generator.

\emph{Case 4 (Without F2C generator): }In this case, the F2C generator and the discriminator $D_C$ are discarded. Then it is a two-step model, where MS and PAN images are firstly fused by the DMG module, and the initial fusion result $\mathbf{F}_d$ is enhanced by the C2F generator. The training loss of the first step is the DMG loss $L_{DMG}$  and that of the second step is the reconstruction loss and adversarial loss related only to C2F generator.

\begin{table}[ht]
	\centering
	\caption{\protect\centering{Ablation Studies}}\label{table_6}
	\setlength{\tabcolsep}{7mm}{
		\begin{tabular}{c|c|c|c}
			\hline \hline
			Cases            & Q4     & SAM    & ERGAS  \\ \hline
			Baseline         & $\bm{0.9773}$ & $\bm{2.8485}$ & $\bm{2.5370}$ \\ \hline
			Case 1(w/o DMG)  & 0.9346 & 3.4412 & 3.6256 \\ \hline
			Case 2(w/o SSRC) & 0.8549 & 3.8968 & 4.5109 \\ \hline
			Case 3(w/o C2F)  & 0.8327 & 4.7562 & 6.3010 \\ \hline
			Case 4(w/o F2C)  & 0.8438 & 5.0038 & 6.4154 \\
			\hline \hline
	\end{tabular}}
\end{table}

The evaluation indices of above four cases together with the baseline are demonstrated in Table~\ref{table_6}. By comparing \emph{Baseline} and \emph{Case 1}, we find that the SSRC module can help to obtain a basically satisfactory final result even if the initial fusion result is obtained by a simple average operation. It can be observed from the comparison between \emph{Baseline} and \emph{Case 2} that the SSRC module is very important and the better fusion results benefit not only from the SSRC structure but also the adversarial loss and the cycle-consistent loss. \emph{Case 3} and \emph{Case 4}  are used to verify the necessity of two GANs in SSRC module. Once we discard one of them, the cyclic-structure is destroyed and the cyclic-consistent loss is useless. The model in \emph{Case 3} is a bidirectional model and that in \emph{Case 4} is also a two-step model, but both of them cannot obtain well fusion results. These studies demonstrate that each module in PC-GANs is important, and they work together to produce satisfactory results.

\begin{figure}[h]
	\centering
		\subfigure[]{\includegraphics[width=0.155\textwidth]{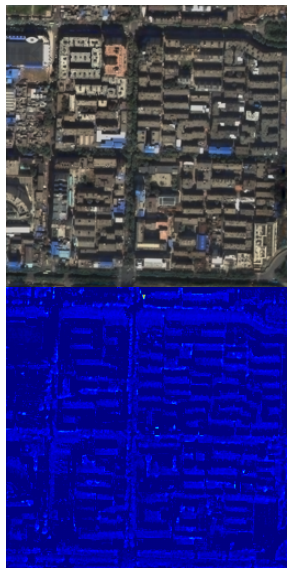}}
		\subfigure[]{\includegraphics[width=0.155\textwidth]{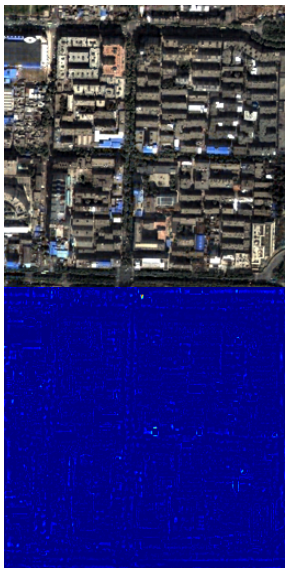}}
		\subfigure[]{\includegraphics[width=0.155\textwidth]{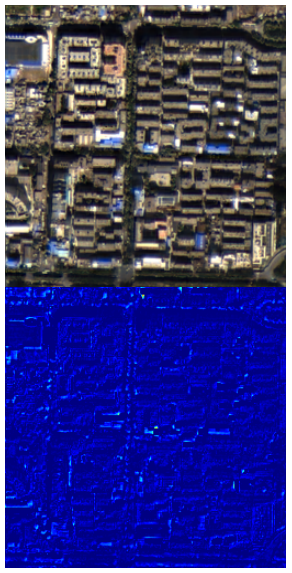}}
		\subfigure[]{\includegraphics[width=0.155\textwidth]{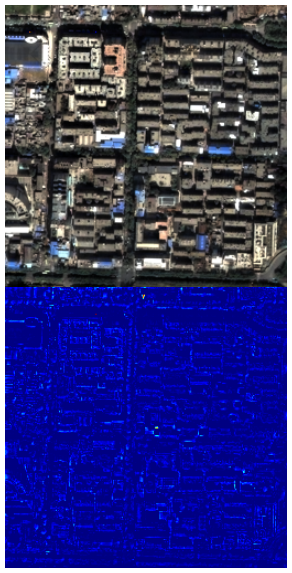}}
		\subfigure[]{\includegraphics[width=0.155\textwidth]{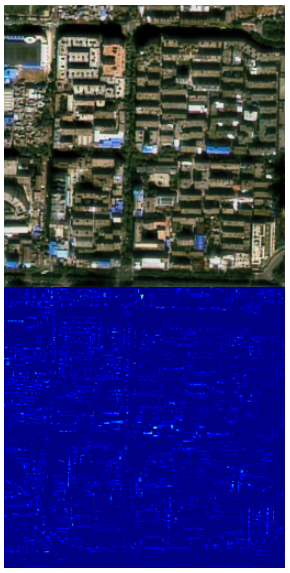}}
		\subfigure[]{\includegraphics[width=0.155\textwidth]{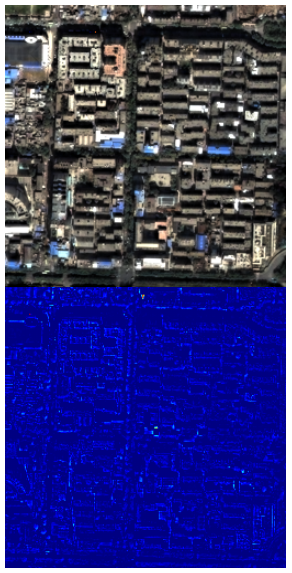}}
		\subfigure[]{\includegraphics[width=0.155\textwidth]{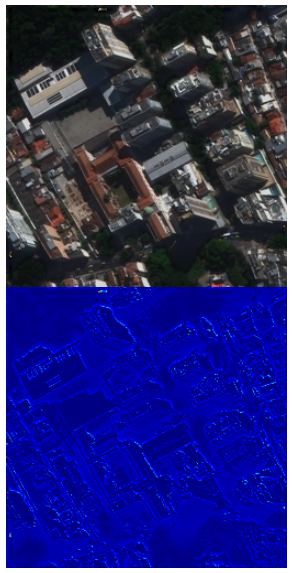}}
		\subfigure[]{\includegraphics[width=0.155\textwidth]{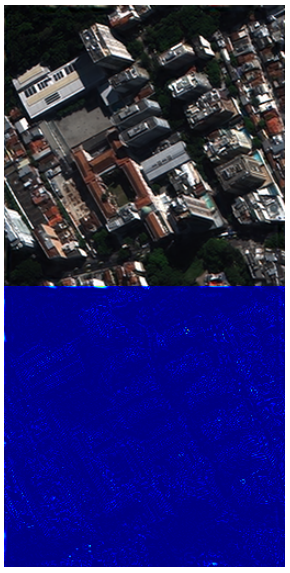}}
		\subfigure[]{\includegraphics[width=0.155\textwidth]{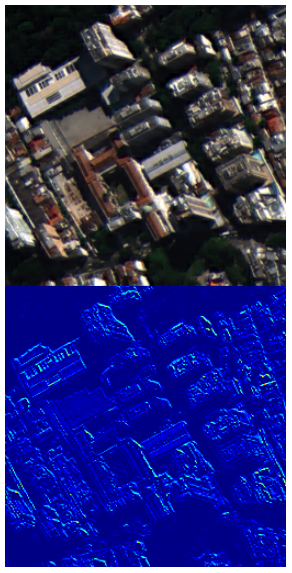}}
		\subfigure[]{\includegraphics[width=0.155\textwidth]{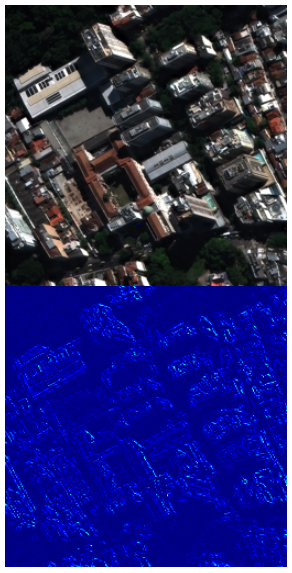}}
		\subfigure[]{\includegraphics[width=0.155\textwidth]{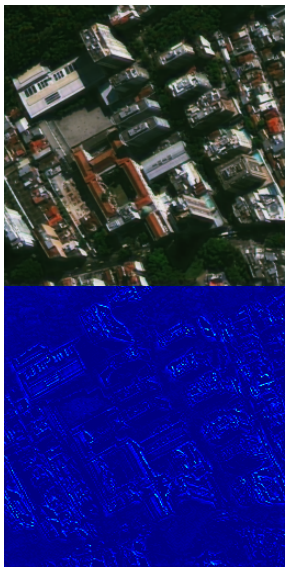}}
		\subfigure[]{\includegraphics[width=0.155\textwidth]{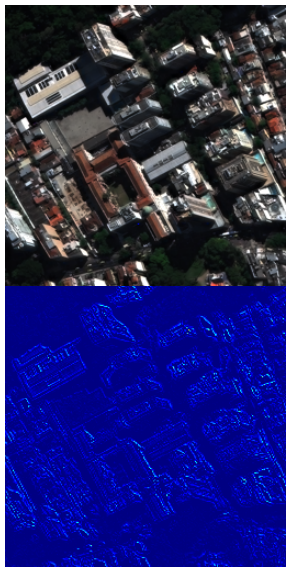}}
		\subfigure{\includegraphics[width=0.5\textwidth]{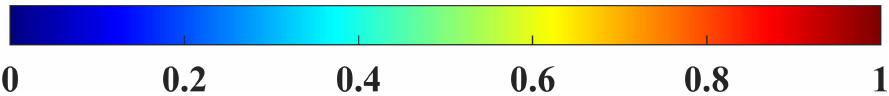}}
	  \caption{Comparison experiments on the progressive compensation strategy. The fusion results and error maps of (a) DMG, (b) DMG-SSRC (PC-GANs), (c) GTP-PNet \cite{zhang2021gtp}, (d) GTP-PNet-SSRC, (e) PanGAN \cite{ma2020pan}, and (f) PanGAN-SSRC on QuickBird data. The fusion results and error maps of (g) DMG, (h) DMG-SSRC (PC-GANs), (i) GTP-PNet \cite{zhang2021gtp}, (j) GTP-PNet-SSRC, (k) PanGAN \cite{ma2020pan}, and (l) PanGAN-SSRC on WorldView-4 data.}\label{fig13}
\end{figure}

\section{Analysis of SSRC Strategy}\label{sec_ASS}

Since the DMG module can be used as a complete network to accomplish the pan-sharpening task, the first experiment is taken on the model that only utilizes the DMG module. Experimental results and quality indexes are shown in Figure \ref{fig13} and Table~\ref{table_7}. Although the use of dilated convolutions will extend the receptive field, images sharpened only by the DMG module are spectrally distorted and are lack of spatial details as well. The reasons are that the relatively deep network will result in information loss when back-propagating the gradients and the directly up-sampling of the coefficients will also introduce some inconsistency. However, results shown in Figure \ref{fig13} (b) and Figure \ref{fig13} (h) have a great improvement. Consequently, the DMG module should also be refined by the SSRC module to realize a progressive compensation of spatial and spectral residuals.

\begin{table}[ht]
	\centering
	\caption{\protect\centering{Investigations on Progressive Compensation Strategy}}\label{table_7}
	
	\begin{tabular}{c|c|c|c|c|c|c}
		\hline \hline
		\multirow{2}{*}{Method} &\multicolumn{2}{c|}{Q4} &\multicolumn{2}{c|}{SAM} &\multicolumn{2}{c}{ERGAS}    \\ \cline{2-7}
		&QB        &WV-4                 &QB              &WV-4     &QB               &WV-4     \\ \hline
		DMG                     &0.8549    &0.8261               &3.8968          &6.6418   &4.5109           &7.1171   \\ \hline
		DMG-SSRC             &$\bm{0.9773}$    &$\bm{0.9361}$               &$\bm{2.8485}$          &$\bm{5.4779}$   &$\bm{2.5370}$           &$\bm{4.8823}$   \\ \hline
		GTP-PNet                 &0.9529    &0.9034				 &3.8366		  &5.8316   &2.8074          &5.1840   \\ \hline
		GTP-PNet-SSRC                  &0.9616    &0.9087               &3.0184          &5.7075   &2.7981           &5.0854   \\ \hline
		PanGAN                &0.9427    &0.9068               &4.5248          &6.6713   &2.9539           &5.3160   \\ \hline
		PanGAN-SSRC                &0.9538    &0.9155               &2.9786          &5.7194   &2.7897           &5.0018   \\
		\hline \hline
	\end{tabular}
\end{table}

One unanticipated finding is that even though the progressive compensation strategy is designed for DMG network, the trained SSRC structure can also be used to refine fusion results of other methods. To investigate the generalization of progressive compensation strategy to other pan-sharpening algorithms, we take GTP-PNet \cite{zhang2021gtp} and PanGAN \cite{ma2020pan} results as the inputs of SSRC structure respectively to obtain GTP-PNet-SSRC and PanGAN-SSRC results. Figure \ref{fig13} also shows the results and corresponding error maps for this experiment. From the error maps and the indexes shown in Table~\ref{table_7}, we can observe that the refined results are better than original results in the spatial and spectral domain. Though the error maps of PanGAN [32] and PanGAN-SSRC are not evident, the spectral information has been corrected by the SSRC. The reason why GTP-PNet-SSRC and PanGAN-SSRC results are relatively inferior to DMG-SSRC (PC-GANs) is that in DMG-SSRC, the SSRC module not only refines the DMG result but also assists the DMG module in updating its parameters.

Above experiments demonstrate that the progressive compensation strategy presented in this paper has generalization ability and can be transplanted as a post-processing operation to other pan-sharpening methods.

\section{Conclusions}\label{sec_CON}
We present a PC-GANs model, which focuses on multisource image fusion and is used for pan-sharpening in this paper. In the proposed PC-GANs based pan-sharpening method, the MS image is enhanced in a two-step scheme. In the first step, a deep multiscale guidance network is used to pan-sharpen the MS image, and in the second step, the fusion result is further enhanced by progressively compensating the spatial-spectral residuals, which is realized by the SSRC structure that is composed of a couple of multiscale GANs who have reversed structures. As a result, triple GANs constitute the PC-GANs model. Based on such a specific structure, we design a joint compensation loss function that not only considers the reconstruction loss but also takes the adversarial loss and cycle-consistent loss into account. Furthermore, the progressive compensation strategy can also be used for other pan-sharpening methods. Experimental results on datasets from QB and WV-4 satellites show the effectiveness of the proposed PC-GANs model.

\bibliography{0_PC_GAN}

\begin{thebibliography}{10}
\expandafter\ifx\csname url\endcsname\relax
  \def\url#1{\texttt{#1}}\fi
\expandafter\ifx\csname urlprefix\endcsname\relax\def\urlprefix{URL }\fi
\expandafter\ifx\csname href\endcsname\relax
  \def\href#1#2{#2} \def\path#1{#1}\fi

\bibitem{shu2015object}
Y.~Shu, H.~Tang, J.~Li, T.~Mao, S.~He, A.~Gong, Y.~Chen, H.~Du, Object-based
  unsupervised classification of vhr panchromatic satellite images by combining
  the hdp and ibp on multiple scenes, IEEE Transactions on Geoscience and
  Remote Sensing 53~(11) (2015) 6148--6162.

\bibitem{yang2018self}
S.~Yang, Z.~Feng, M.~Wang, K.~Zhang, Self-paced learning-based probability
  subspace projection for hyperspectral image classification, IEEE transactions
  on neural networks and learning systems 30~(2) (2018) 630--635.

\bibitem{liu2016deep}
J.~Liu, M.~Gong, K.~Qin, P.~Zhang, A deep convolutional coupling network for
  change detection based on heterogeneous optical and radar images, IEEE
  transactions on neural networks and learning systems 29~(3) (2016) 545--559.

\bibitem{zhang2018coarse}
W.~Zhang, X.~Lu, X.~Li, A coarse-to-fine semi-supervised change detection for
  multispectral images, IEEE Transactions on Geoscience and Remote Sensing
  56~(6) (2018) 3587--3599.

\bibitem{ghamisi2013multilevel}
P.~Ghamisi, M.~S. Couceiro, F.~M. Martins, J.~A. Benediktsson, Multilevel image
  segmentation based on fractional-order darwinian particle swarm optimization,
  IEEE Transactions on Geoscience and Remote sensing 52~(5) (2013) 2382--2394.

\bibitem{vivone2014critical}
G.~Vivone, L.~Alparone, J.~Chanussot, M.~Dalla~Mura, A.~Garzelli, G.~A.
  Licciardi, R.~Restaino, L.~Wald, A critical comparison among pansharpening
  algorithms, IEEE Transactions on Geoscience and Remote Sensing 53~(5) (2014)
  2565--2586.

\bibitem{baronti2011theoretical}
S.~Baronti, B.~Aiazzi, M.~Selva, A.~Garzelli, L.~Alparone, A theoretical
  analysis of the effects of aliasing and misregistration on pansharpened
  imagery, IEEE Journal of Selected Topics in Signal Processing 5~(3) (2011)
  446--453.

\bibitem{yin2017joint}
H.~Yin, A joint sparse and low-rank decomposition for pansharpening of
  multispectral images, IEEE transactions on geoscience and remote sensing
  55~(6) (2017) 3545--3557.

\bibitem{leung2013improved}
Y.~Leung, J.~Liu, J.~Zhang, An improved adaptive intensity--hue--saturation
  method for the fusion of remote sensing images, IEEE Geoscience and Remote
  Sensing Letters 11~(5) (2013) 985--989.

\bibitem{ghahremani2016nonlinear}
M.~Ghahremani, H.~Ghassemian, Nonlinear ihs: A promising method for
  pan-sharpening, IEEE Geoscience and Remote Sensing Letters 13~(11) (2016)
  1606--1610.

\bibitem{shahdoosti2017pansharpening}
H.~R. Shahdoosti, N.~Javaheri, Pansharpening of clustered ms and pan images
  considering mixed pixels, IEEE Geoscience and Remote Sensing Letters 14~(6)
  (2017) 826--830.

\bibitem{yang2010image}
S.~Yang, M.~Wang, L.~Jiao, R.~Wu, Z.~Wang, Image fusion based on a new
  contourlet packet, Information Fusion 11~(2) (2010) 78--84.

\bibitem{nencini2007remote}
F.~Nencini, A.~Garzelli, S.~Baronti, L.~Alparone, Remote sensing image fusion
  using the curvelet transform, Information fusion 8~(2) (2007) 143--156.

\bibitem{xing2018pansharpening}
Y.~Xing, M.~Wang, S.~Yang, K.~Zhang, Pansharpening with multiscale geometric
  support tensor machine, IEEE Transactions on Geoscience and Remote Sensing
  56~(5) (2018) 2503--2517.

\bibitem{restaino2016fusion}
R.~Restaino, G.~Vivone, M.~Dalla~Mura, J.~Chanussot, Fusion of multispectral
  and panchromatic images based on morphological operators, IEEE Transactions
  on Image Processing 25~(6) (2016) 2882--2895.

\bibitem{vivone2014pansharpening}
G.~Vivone, M.~Sim{\~o}es, M.~Dalla~Mura, R.~Restaino, J.~M. Bioucas-Dias, G.~A.
  Licciardi, J.~Chanussot, Pansharpening based on semiblind deconvolution, IEEE
  Transactions on Geoscience and Remote Sensing 53~(4) (2014) 1997--2010.

\bibitem{he2014new}
X.~He, L.~Condat, J.~M. Bioucas-Dias, J.~Chanussot, J.~Xia, A new pansharpening
  method based on spatial and spectral sparsity priors, IEEE Transactions on
  Image Processing 23~(9) (2014) 4160--4174.

\bibitem{huang2015new}
W.~Huang, L.~Xiao, Z.~Wei, H.~Liu, S.~Tang, A new pan-sharpening method with
  deep neural networks, IEEE Geoscience and Remote Sensing Letters 12~(5)
  (2015) 1037--1041.

\bibitem{xing2018pan}
Y.~Xing, M.~Wang, S.~Yang, L.~Jiao, Pan-sharpening via deep metric learning,
  ISPRS Journal of Photogrammetry and Remote Sensing 145 (2018) 165--183.

\bibitem{masi2016pansharpening}
G.~Masi, D.~Cozzolino, L.~Verdoliva, G.~Scarpa, Pansharpening by convolutional
  neural networks, Remote Sensing 8~(7) (2016) 594.

\bibitem{wei2017boosting}
Y.~Wei, Q.~Yuan, H.~Shen, L.~Zhang, Boosting the accuracy of multispectral
  image pansharpening by learning a deep residual network, IEEE Geoscience and
  Remote Sensing Letters 14~(10) (2017) 1795--1799.

\bibitem{yang2017pannet}
J.~Yang, X.~Fu, Y.~Hu, Y.~Huang, X.~Ding, J.~Paisley, Pannet: A deep network
  architecture for pan-sharpening, in: Proceedings of the IEEE international
  conference on computer vision, 2017, pp. 5449--5457.

\bibitem{yuan2018multiscale}
Q.~Yuan, Y.~Wei, X.~Meng, H.~Shen, L.~Zhang, A multiscale and multidepth
  convolutional neural network for remote sensing imagery pan-sharpening, IEEE
  Journal of Selected Topics in Applied Earth Observations and Remote Sensing
  11~(3) (2018) 978--989.

\bibitem{dong2015image}
C.~Dong, C.~C. Loy, K.~He, X.~Tang, Image super-resolution using deep
  convolutional networks, IEEE transactions on pattern analysis and machine
  intelligence 38~(2) (2015) 295--307.

\bibitem{scarpa2018target}
G.~Scarpa, S.~Vitale, D.~Cozzolino, Target-adaptive cnn-based pansharpening,
  IEEE Transactions on Geoscience and Remote Sensing 56~(9) (2018) 5443--5457.

\bibitem{ledig2017photo}
C.~Ledig, L.~Theis, F.~Husz{\'a}r, J.~Caballero, A.~Cunningham, A.~Acosta,
  A.~Aitken, A.~Tejani, J.~Totz, Z.~Wang, et~al., Photo-realistic single image
  super-resolution using a generative adversarial network, in: Proceedings of
  the IEEE conference on computer vision and pattern recognition, 2017, pp.
  4681--4690.

\bibitem{ma2020pan}
J.~Ma, W.~Yu, C.~Chen, P.~Liang, X.~Guo, J.~Jiang, Pan-gan: An unsupervised
  pan-sharpening method for remote sensing image fusion, Information Fusion 62
  (2020) 110--120.

\bibitem{liu2020psgan}
Q.~Liu, H.~Zhou, Q.~Xu, X.~Liu, Y.~Wang, Psgan: A generative adversarial
  network for remote sensing image pan-sharpening, IEEE Transactions on
  Geoscience and Remote Sensing.

\bibitem{zhu2020super}
X.~Zhu, Y.~Cheng, J.~Peng, R.~Wang, M.~Le, X.~Liu, Super-resolved image
  perceptual quality improvement via multifeature discriminators, Journal of
  Electronic Imaging 29~(1) (2020) 013017.

\bibitem{gastineau2021generative}
A.~Gastineau, J.-F. Aujol, Y.~Berthoumieu, C.~Germain, Generative adversarial
  network for pansharpening with spectral and spatial discriminators, IEEE
  Transactions on Geoscience and Remote Sensing.

\bibitem{he2012guided}
K.~He, J.~Sun, X.~Tang, Guided image filtering, IEEE transactions on pattern
  analysis and machine intelligence 35~(6) (2012) 1397--1409.

\bibitem{hastie2009elements}
T.~Hastie, R.~Tibshirani, J.~Friedman, The elements of statistical learnin,
  Cited on (2009) 33.

\bibitem{chen2017fast}
Q.~Chen, J.~Xu, V.~Koltun, Fast image processing with fully-convolutional
  networks, in: Proceedings of the IEEE International Conference on Computer
  Vision, 2017, pp. 2497--2506.

\bibitem{mao2017least}
X.~Mao, Q.~Li, H.~Xie, R.~Y. Lau, Z.~Wang, S.~Paul~Smolley, Least squares
  generative adversarial networks, in: Proceedings of the IEEE international
  conference on computer vision, 2017, pp. 2794--2802.

\bibitem{zhu2017unpaired}
J.-Y. Zhu, T.~Park, P.~Isola, A.~A. Efros, Unpaired image-to-image translation
  using cycle-consistent adversarial networks, in: Proceedings of the IEEE
  international conference on computer vision, 2017, pp. 2223--2232.

\bibitem{aiazzi2007improving}
B.~Aiazzi, S.~Baronti, M.~Selva, Improving component substitution pansharpening
  through multivariate regression of ms $+ $ pan data, IEEE Transactions on
  Geoscience and Remote Sensing 45~(10) (2007) 3230--3239.

\bibitem{aiazzi2003mtf}
B.~Aiazzi, L.~Alparone, S.~Baronti, A.~Garzelli, M.~Selva, An mtf-based
  spectral distortion minimizing model for pan-sharpening of very high
  resolution multispectral images of urban areas, in: 2003 2nd GRSS/ISPRS Joint
  Workshop on Remote Sensing and Data Fusion over Urban Areas, IEEE, 2003, pp.
  90--94.

\bibitem{jiang2013two}
C.~Jiang, H.~Zhang, H.~Shen, L.~Zhang, Two-step sparse coding for the
  pan-sharpening of remote sensing images, IEEE journal of selected topics in
  applied earth observations and remote sensing 7~(5) (2013) 1792--1805.

\bibitem{zhang2021gtp}
H.~Zhang, J.~Ma, Gtp-pnet: A residual learning network based on gradient
  transformation prior for pansharpening, ISPRS Journal of Photogrammetry and
  Remote Sensing 172 (2021) 223--239.

\bibitem{wald1997fusion}
L.~Wald, T.~Ranchin, M.~Mangolini, Fusion of satellite images of different
  spatial resolutions: Assessing the quality of resulting images,
  Photogrammetric engineering and remote sensing 63~(6) (1997) 691--699.

\bibitem{alparone2008multispectral}
L.~Alparone, B.~Aiazzi, S.~Baronti, A.~Garzelli, F.~Nencini, M.~Selva,
  Multispectral and panchromatic data fusion assessment without reference,
  Photogrammetric Engineering \& Remote Sensing 74~(2) (2008) 193--200.

\end{thebibliography}

\end{document}